\documentclass{nature}

\usepackage[top=0.8in, bottom=1in, left=0.8in, right=0.8in]{geometry}

\usepackage{graphicx}%
\usepackage{multirow}%
\usepackage{amsmath,amssymb,amsfonts}%
\usepackage{amsthm}%
\usepackage{mathrsfs}%
\usepackage[title]{appendix}%
\usepackage{xcolor}%
\usepackage{textcomp}%
\usepackage{manyfoot}%
\usepackage{booktabs}%
\usepackage{algorithm}%
\usepackage{algorithmicx}%
\usepackage{algpseudocode}%
\usepackage{listings}%
\usepackage{ulem}
\usepackage{hyperref}
\usepackage{caption} 
\usepackage[OT1]{fontenc}

\DeclareCaptionLabelFormat{nospace}{#1#2}
\newcommand{\sersic}{S${\rm \acute{e}}$rsic}

\title{In-Situ Spheroid Formation in Distant Submillimeter-Bright Galaxies}

\author{\large Qing-Hua Tan$^{1,2}$,
Emanuele Daddi$^{2}$,
Benjamin Magnelli$^{2}$,
Camila A. Correa$^{2}$,
Fr\'{e}d\'{e}ric Bournaud$^{2}$,
Sylvia Adscheid$^{3}$,
Shao-Bo Zhang$^{1}$,
David Elbaz$^{2}$,
Carlos G\'{o}mez-Guijarro$^{2}$,
Boris S. Kalita$^{4,5,6}$,
Daizhong Liu$^{1}$,
Zhaoxuan Liu$^{4,5,7}$,
J\'{e}r\^{o}me Pety$^{8,9}$,
Annagrazia Puglisi$^{10,11}$,
Eva Schinnerer$^{12}$,
John D. Silverman$^{4,5,7,13}$,
Francesco Valentino$^{14,15}$
\vspace{8pt}}

\begin{document}

\maketitle

\begin{affiliations}
\small
 \item  Purple Mountain Observatory, Chinese Academy of Sciences, 10 Yuanhua Road, Nanjing 210023, People{\textquotesingle}s Republic of China; \href{mailto: qhtan@pmo.ac.cn}{\nolinkurl{qhtan@pmo.ac.cn}}
 \item  Universit\'{e} Paris-Saclay, Universit\'{e} Paris Cit\'{e}, CEA, CNRS, AIM,  Gif-sur-Yvette 91191, France; \href{mailto: edaddi@cea.fr}{\nolinkurl{edaddi@cea.fr}}
 \item Argelander-Institut f\"{u}r Astronomie, Universit\"{a}t Bonn, Auf dem H\"{u}gel 71, 53121 Bonn, Germany
 \item Kavli Institute for the Physics and Mathematics of the Universe, The University of Tokyo, Kashiwa, 277-8583, Japan
 \item Center for Data-Driven Discovery, Kavli IPMU (WPI), UTIAS, The University of Tokyo, Kashiwa, Chiba 277-8583, Japan
 \item Kavli Institute for Astronomy and Astrophysics, Peking University, Beijing 100871, People{\textquotesingle}s Republic of China
 \item Department of Astronomy, School of Science, The University of Tokyo, 7-3-1 Hongo, Bunkyo, Tokyo 113-0033, Japan
 \item Institut de Radioastronomie Millim\'{e}trique, 300 Rue de la Piscine, 38406 Saint-Martin d'H\`{e}res, France
 \item LERMA, Observatoire de Paris, PSL Research University, CNRS, Sorbonne Universit\'es, 75014 Paris, France
 \item School of Physics and Astronomy, University of Southampton, Highfield SO17 1BJ, UK
 \item Center for Extragalactic Astronomy, Department of Physics, Durham University, South Road, Durham DH1 3LE, UK
 \item Max-Planck-Institut f\"{u}r Astronomie, K\"{o}nigstuhl 17, 69117 Heidelberg, Germany
 \item Center for Astrophysical Sciences, Department of Physics \& Astronomy, Johns Hopkins University, Baltimore, MD 21218, USA
 \item European Southern Observatory, Karl-Schwarzschild-Str. 2, D-85748 Garching bei Munchen, Germany
 \item Cosmic Dawn Center (DAWN), Denmark
\end{affiliations}

\begin{abstract}
\textbf{The majority of stars in today's Universe reside within spheroids, which are bulges of spiral galaxies and elliptical galaxies~\cite{Baldry2006, Gadotti2009}. 
Their formation is still an unsolved problem~\cite{Brooks2016,Oser2012,Peng2010}. Infrared/submm-bright galaxies at high redshifts~\cite{Blain2002} have long been suspected to be related to spheroids formation~\cite{Lilly1999,Archibald2002, Dunne2003,DeLucia2006,Tacconi2008,Brisbin2017}. Proving this connection has been hampered so far by heavy dust obscuration when focusing on their stellar emission~\cite{LeBail2023,Cardona-Torres2023,Gillman2023} or by methodologies and limited signal-to-noise ratios when looking at submm wavelengths~\cite{Gullberg2019,Hodge2019}. Here we show that spheroids are directly generated by star formation within the cores of highly luminous starburst galaxies in the distant Universe. This follows from the ALMA submillimeter surface brightness profiles which deviate significantly from those of exponential disks, and from the skewed-high axis-ratio distribution. The majority of these galaxies are fully triaxial rather than flat disks: the ratio of the shortest to the longest of their three axes is half, on average, and increases with spatial compactness. These observations, supported by simulations, reveal a cosmologically relevant pathway for in-situ spheroid formation through starbursts likely preferentially triggered by interactions (and mergers) acting on galaxies fed by non-co-planar gas accretion streams. }
\end{abstract}

\begin{figure*}[htbp]%
\centering
\includegraphics[width=0.4\linewidth]{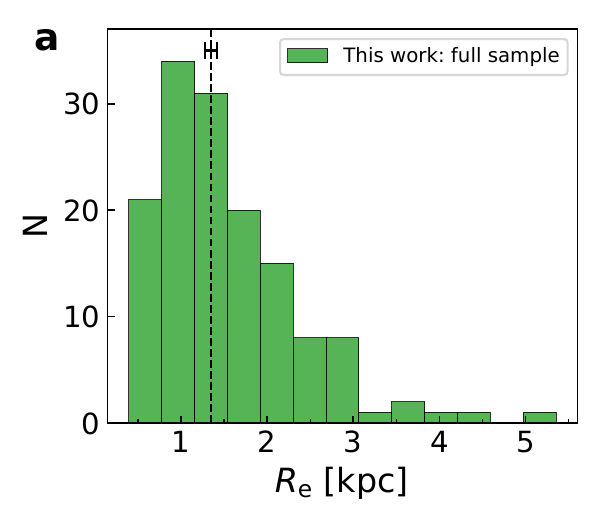}
\hspace{-10pt}
\includegraphics[width=0.4\linewidth]{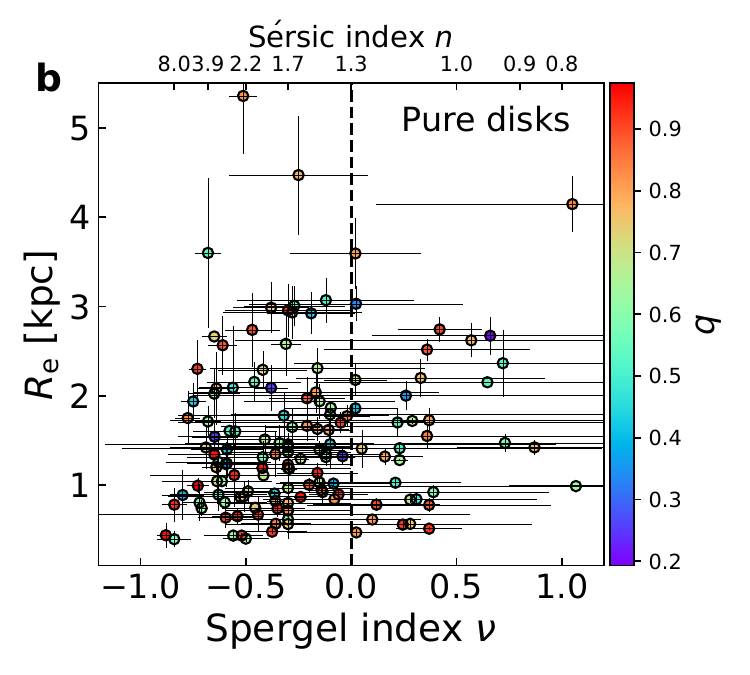}
\includegraphics[width=0.4\linewidth]{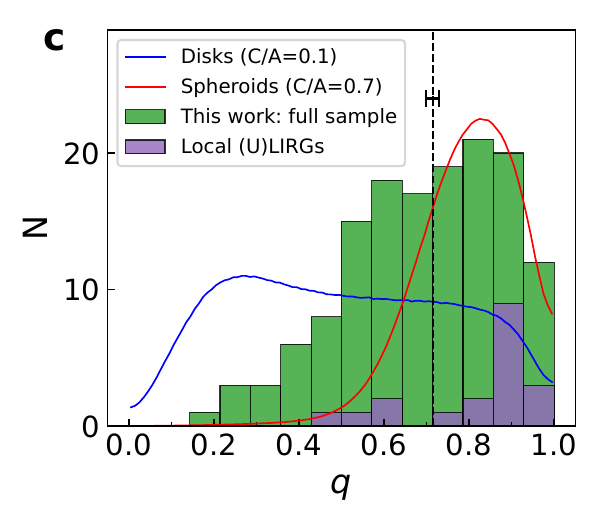}
\hspace{-10pt}
\includegraphics[width=0.42\linewidth]{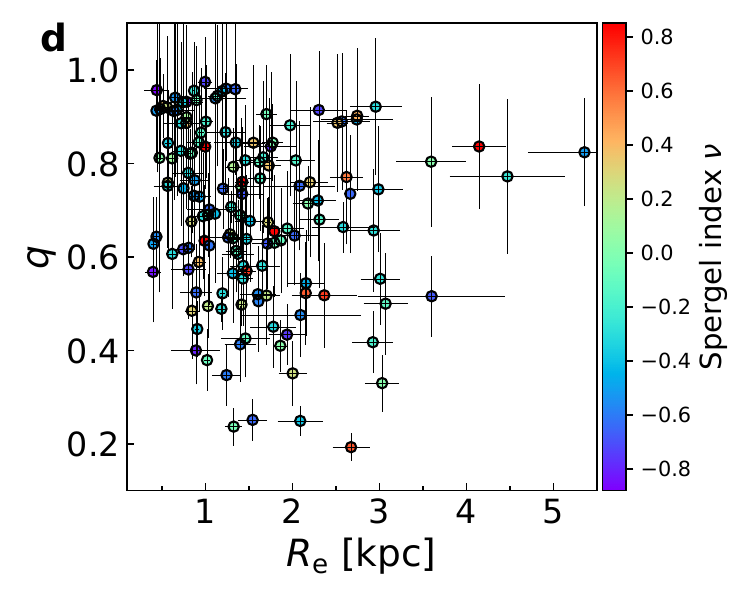}
\includegraphics[width=0.4\linewidth]{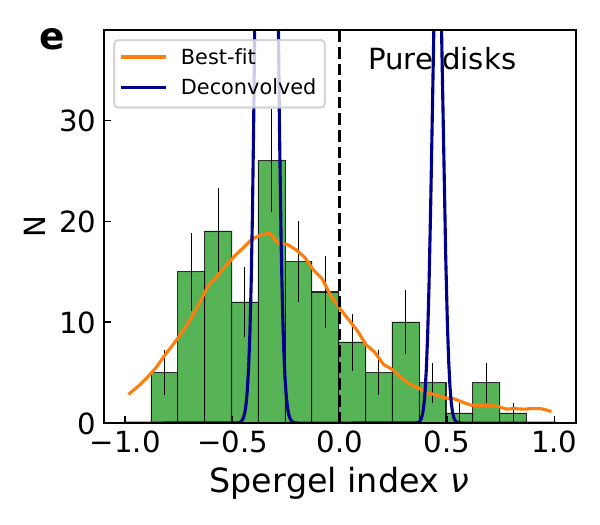}
\hspace{-10pt}
\includegraphics[width=0.4\linewidth]{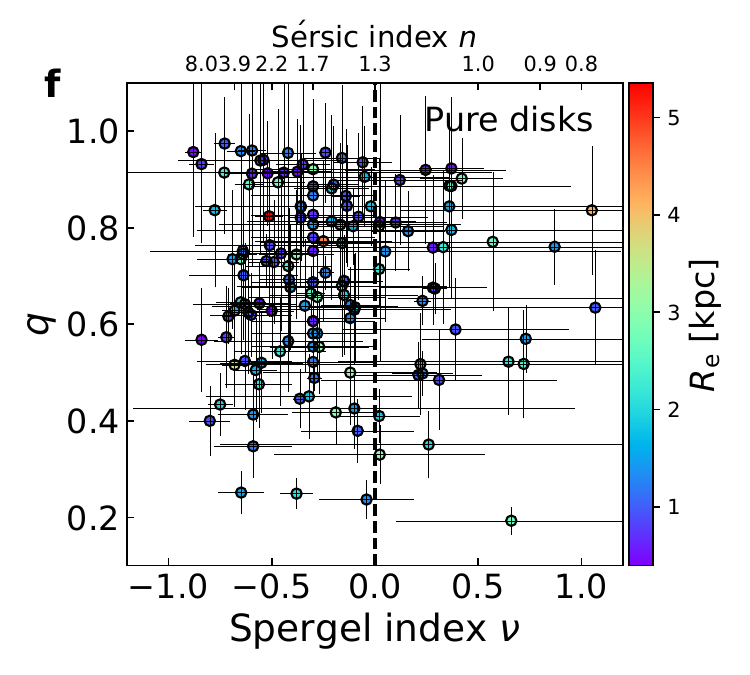}
\caption{\textbf{Physical properties of submm-bright galaxies,  measured from fitting Spergel light profiles to the visibilities.} \textbf{Left}: distributions of measured parameters. \textbf{a}, half-light radius $R_{\rm e}$. \textbf{c}, comparison of the distributions of the observed axis ratio \textit{q} between our sample of galaxies (green) and local (U)LIRGs (purple). As reference, we plot the expected \textit{q} distributions for disks with a scale-height (C/A) of 0.1 (blue curve) and spheroids with a scale-height of 0.7 (red curve). \textbf{e}, Spergel index $\nu$. The blue curves in panel \textbf{e} represent the intrinsic distribution of $\nu$, which was modeled using a combination of two Gaussian functions, while the orange curve represents the best-fit $\nu$ distribution of sample galaxies by taking into account the uncertainty of $\nu$ measurement $\sigma_\nu$ (see Methods). The error bar in each bin corresponds to the 1$\sigma$ Poisson error.  The vertical dashed lines in panels \textbf{a,c} represent the median value of the distribution and the horizontal bar indicates the error on the median, respectively. \textbf{Right}: relationships between measured parameters. \textbf{b},  $R_{\rm e}$ versus Spergel index $\nu$. \textbf{d}, \textit{q} versus $R_{\rm e}$. \textbf{f}, \textit{q} versus $\nu$. \textbf{b,d,f}, data points are colour coded by \textit{q}, $\nu$, and $R_{\rm e}$, respectively. The dashed lines in panels \textbf{b,e,f} mark the  $\nu=0$ threshold above which we classify galaxies as pure disks. \textbf{b,f}, the \sersic \ indices listed on the top axis were converted using Eq.~(\ref{eq:conversion}) (see Methods),  assuming $R_{\rm e}/\theta_{\rm b}=0.41$, where $\theta_{\rm b}$ is the synthesized circularized beam size of the observations. This value is the median measured for the whole sample of galaxies.}\label{fig:histogram}
\vspace{-10pt}
\end{figure*}

We harvested deep ALMA archival observations at 240/340~GHz in the COSMOS and GOODS-S fields~\cite{Adscheid2024}, yielding a submillimeter galaxy (SMG) sample of 146 galaxies that are bright in submm emission (median $S_{\rm 870 \mu m}$ of 7.8~mJy) using the sole criterion of having a ratio of total source flux to the noise [per beam]  (S/N$_{\rm beam}$)~$>$~50 (median S/N$_{\rm beam}$ of 68). 
The high S/N$_{\rm beam}$ threshold is required to enable reliable measurements of morphological parameters~\cite{vanderwel2014a,Magnelli2023}. We quantify the morphology of these galaxies performing fits of bidimensional Spergel~\cite{Spergel2010,Tan2024} functions directly constrained by the interferometric observations  (antenna correlations in the \textit{uv}-plane; Methods), deriving key parameters like Spergel index $\nu$ (the surface brightness profile slope, similar to the \sersic ~\cite{Sersic1968} index; Methods),  half-light radius size $R_{\rm e}$, and axis ratio \textit{q}($\equiv b/a$; where $b$ and $a$ are minor and major axis, respectively), measured with median accuracy of 36\%, 18\%, and 22\%, respectively, based on simulations~\cite{Tan2024}.
These galaxies are submm-compact, massive starbursts, with median stellar mass log($M_\star$)=$11.0_{-0.3}^{+0.2}\ M_\odot$, star formation rate (SFR) of $624_{-333}^{+429}\ M_\odot$~yr$^{-1}$,   
submm half-light radius $R_{\rm e}=1.4_{-0.5}^{+0.6}$~kpc, and redshift $z=2.70^{+0.69}_{-0.73}$ (uncertainties are interquartile ranges; see Methods and Extended Data Table~\ref{tab:properties}).  
Figure~\ref{fig:histogram} shows the distribution of the measured main parameters and the relationships between these quantities.  

Disk-like galaxies have exponential surface brightness profiles~\cite{Dutton2009, Wang&Lilly2022}, corresponding to a \sersic\ index of $n=1$ or, using Spergel functions,  $\nu=0.5$. The majority of our sample demonstrates negative Spergel indices, with a median  $\nu\sim-0.3$, corresponding to $n\sim1.6-2$ after adjusting for the size-dependent conversion from Spergel to \sersic\ \cite{Tan2024}. Profiles with $n>1.5$ suggest the presence of a bulge-like component. Accounting for realistic errors established using simulations allows for deconvolving the observed distributions, indicating nearly bimodal parent distributions with steep profiles for bulge-like systems ($n\sim2$, $\nu\sim-0.4$) (see Fig.~\ref{fig:histogram}e). Disks constitute only $\sim20$\% of the sample and exhibit larger submillimeter sizes compared to bulge-dominated galaxies (see Methods). These findings suggest that the submillimeter emission in these galaxies typically arises from structures that are already spheroid-like, indicating direct in-situ spheroid formation within these systems.

An imprint of the presence of forming bulges should be reflected in the distribution of axis ratio \textit{q}, with a dearth of edge-on systems. 
Figure~\ref{fig:histogram}\textbf{c} shows the histograms of the observed axis ratio in our sample. The distribution is skewed with a large fraction of galaxies having axis ratios in the range of $q=0.6-1.0$, and a deficit of sources at $q<0.4$. 
The median $q$ for the full sample is $0.71_{-0.14}^{+0.13}$ (uncertainties are interquartile ranges). 
There is a clear trend for more compact objects to be rounder, with $q=0.78_{-0.13}^{+0.13}$ for sources with higher SFR surface density ($\Sigma_{\rm SFR}$) versus $0.64_{-0.12}^{+0.13}$ for sources with lower $\Sigma_{\rm SFR}$ when dividing the sample based on $\Sigma_{\rm SFR}$ (see Methods). The compact galaxies are found to be clustered at larger $q>0.8$, in line with the $q$ distribution predicted for spheroidal galaxies (e.g., ref.~\cite{Zhang2022}; see Fig.~\ref{fig:histogram}c).

\begin{figure*}[htbp]%
\centering
\includegraphics[width=0.35\linewidth]{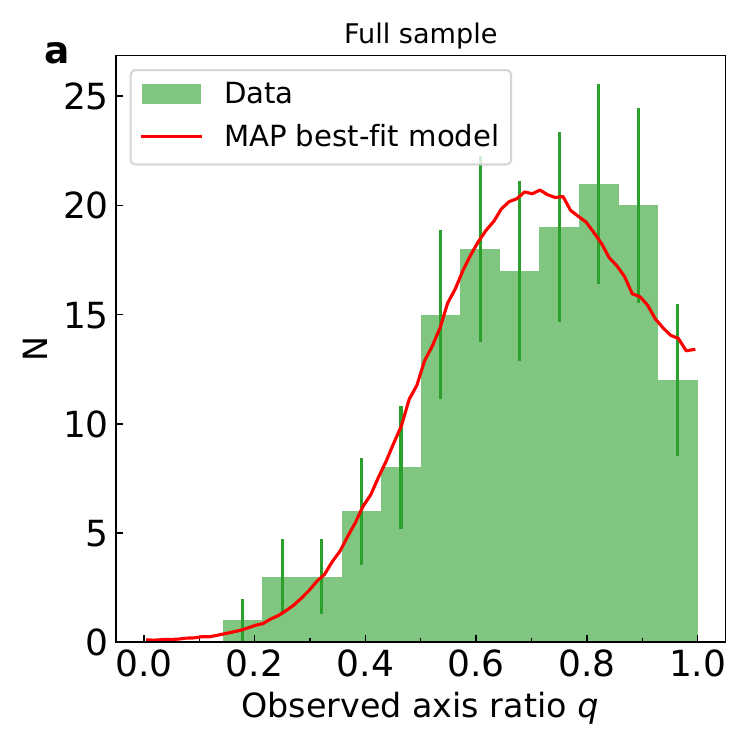}
\includegraphics[width=0.35\linewidth]{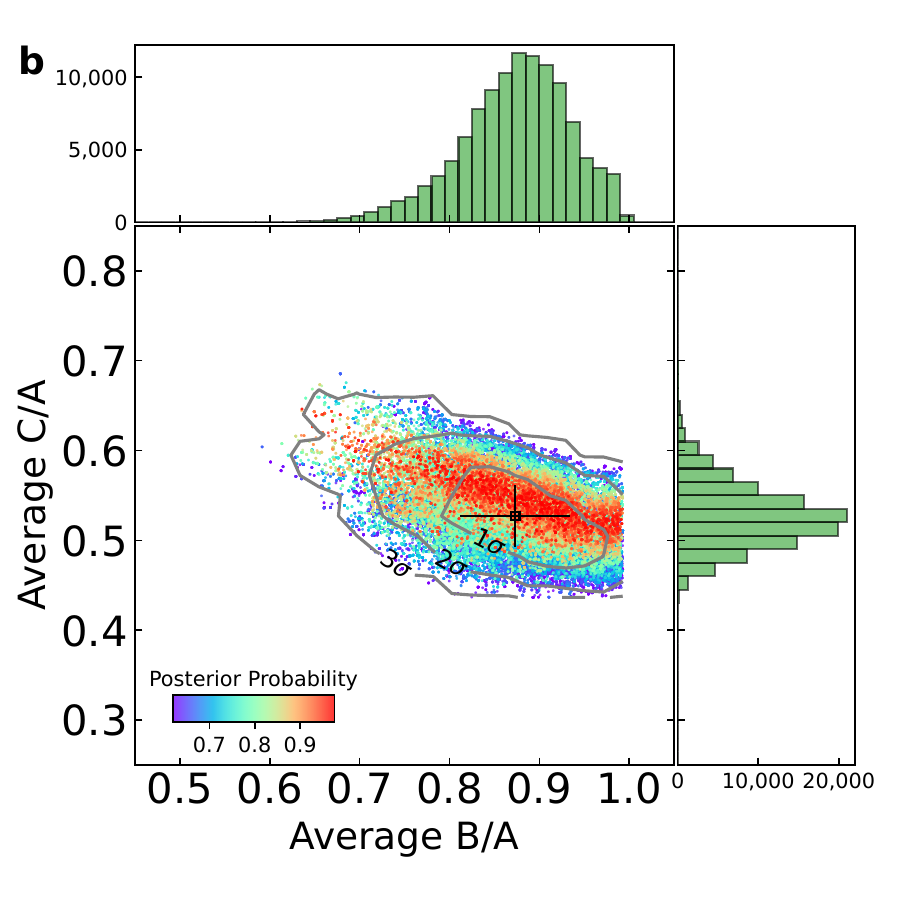}
\includegraphics[width=0.35\linewidth]{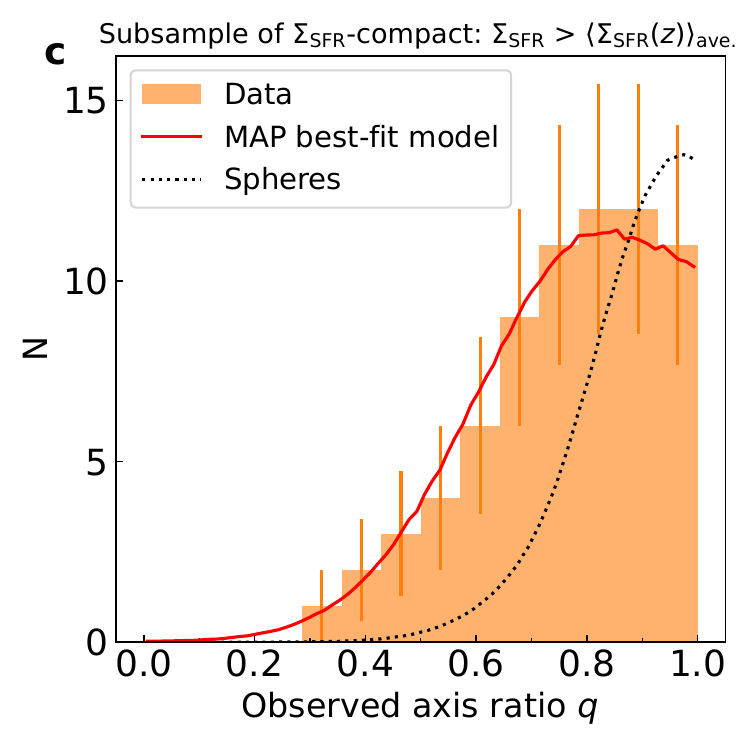}
\includegraphics[width=0.35\linewidth]{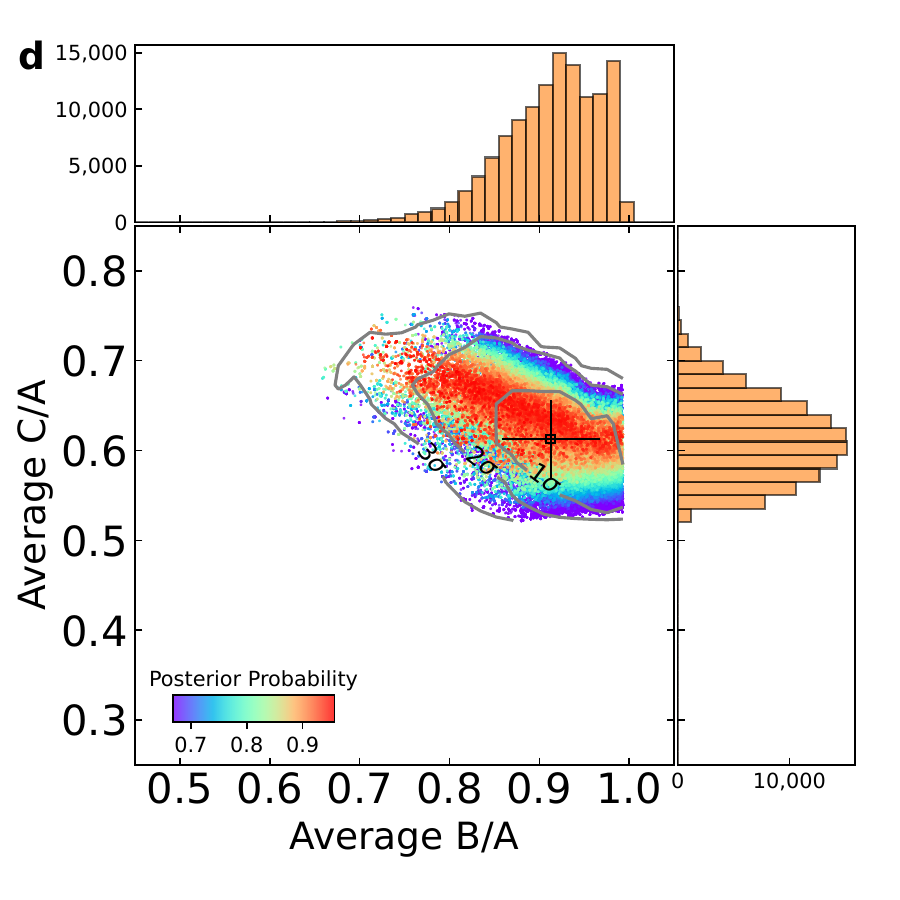}
\includegraphics[width=0.35\linewidth]{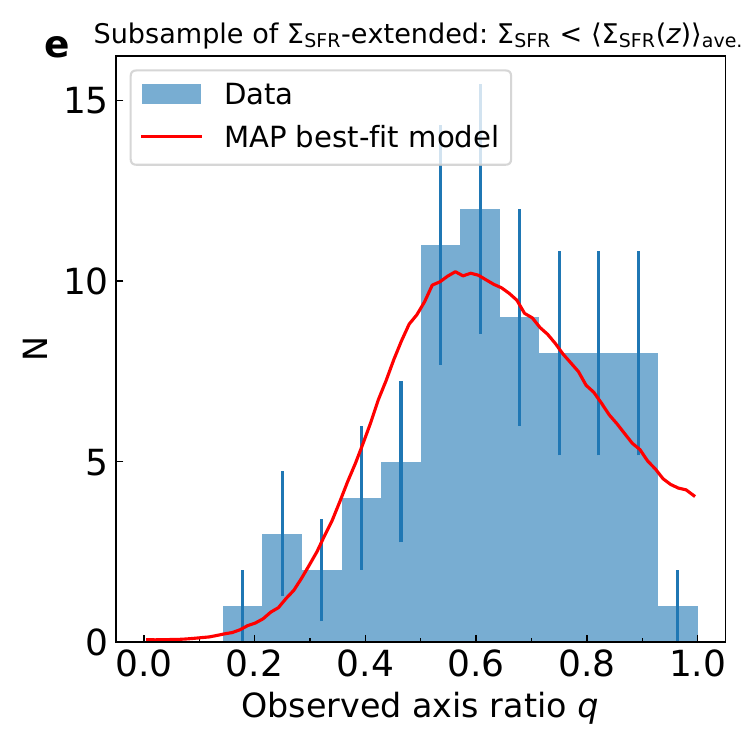}
\includegraphics[width=0.35\linewidth]{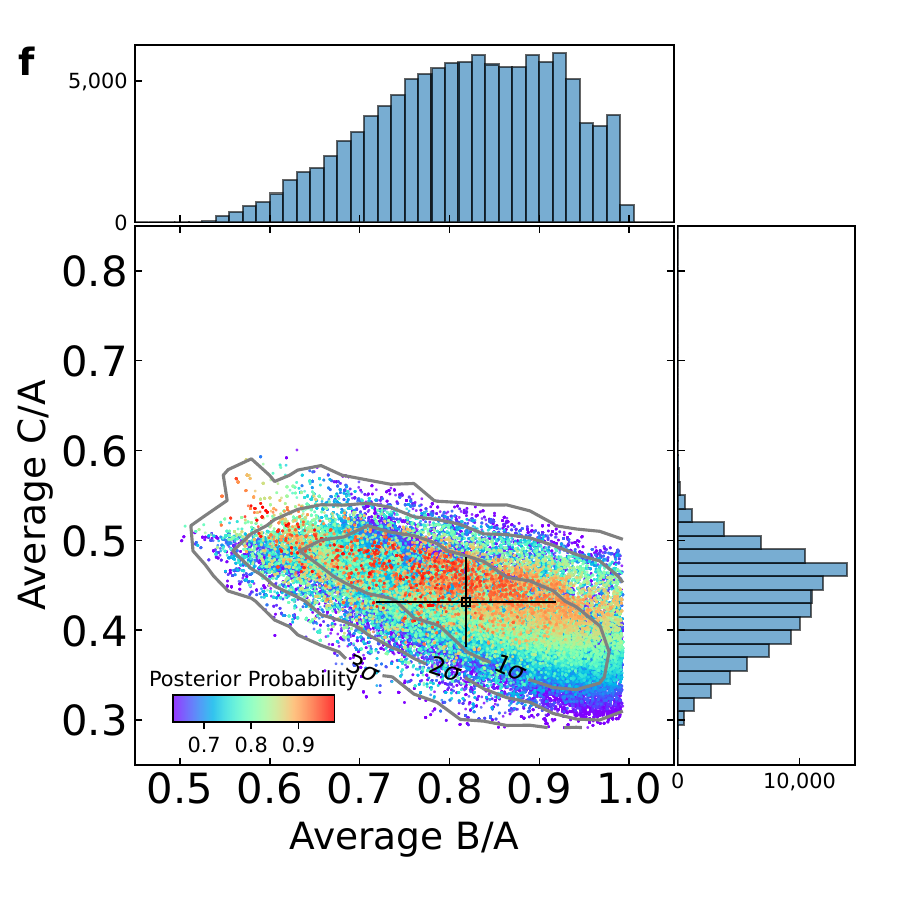}
\caption{\textbf{Observed and 3$D$ intrinsic axis ratios for ALMA submm-bright galaxies.} \textbf{Left}: distribution of observed axis ratio \textit{q}. \textbf{Right}: distribution of average intrinsic face-on axial ratio $B/A$ versus edge-on axial ratio $C/A$ (scale-height), calculated as the first moment of the distribution of B/A and C/A, respectively. The sample distribution of $B/A$ is log-normal, while $C/A$ is Gaussian. \textbf{a,b}: the full sample. \textbf{c,d}: a subsample of $\Sigma_{\rm SFR}$-compact galaxies. \textbf{e,f}: a subsample of $\Sigma_{\rm SFR}$-extended galaxies. The red curves in left panels represent the MAP best-fit model with the highest posterior probability, which is identified as the step with highest posterior probability and nearest to the position defined by averages $\langle B/A \rangle$ and $\langle C/A \rangle$ (the black squares in the right panels; see Methods). The error bar in each histogram bin corresponds to the 1$\sigma$ Poisson error. The dotted curve in panel \textbf{c} represents the expected distribution of projected axis ratios for galaxies with a spherical shape (C/A=1). The data points in right panels are colour coded by the posterior probability. Contours are plotted based on the density of the data points, using the posterior probability as weights. Contour levels are 1$\sigma$, 2$\sigma$, and 3$\sigma$. The black squares and error bars in the right panels represent the weighted mean and standard deviation derived from the histograms. We used the chi-square statistic to quantify the difference between the cumulative distribution function (CDF) of observational data and that of the best-fit model. The full sample and the two subsamples of $\Sigma_{\rm SFR}$-compact and -extended galaxies have $\chi^2 = $3.8, 5.8, and 9.2, respectively.
}
\label{fig:hist-axis}
\vspace{-15pt}
\end{figure*}

\begin{figure}[h]%
\centering
\includegraphics[width=0.95\linewidth]{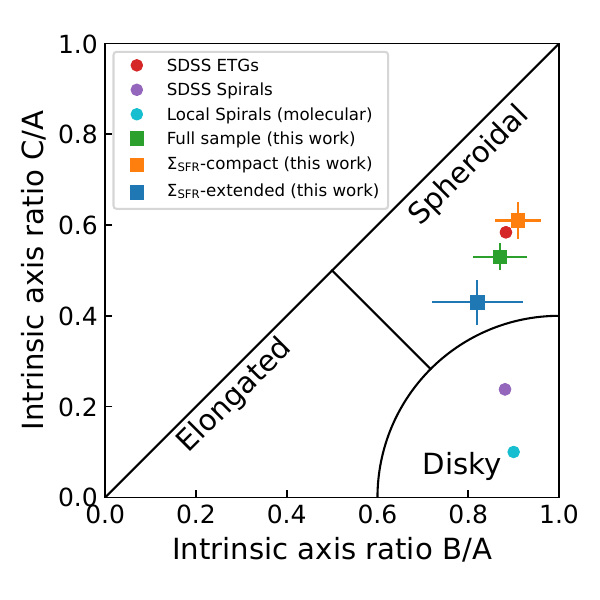}
\caption{\textbf{Classification of intrinsic galaxy shapes.}
The boundaries for disky, elongated, and spheroidal shapes are adopted from ref.~\cite{vanderwel2014b}.
The average shape parameters measured for our sample of galaxies (green square: full sample; orange square: $\Sigma_{\rm SFR}$-compact subsample; blue square: $\Sigma_{\rm SFR}$-extended subsample) are compared to local ETGs (red circle) and late-type galaxies (purple circle) measured in optical emission~\cite{Rodriguez2013}, and a local sample of spirals with 3D molecular gas distribution constrained by modeling~\cite{Patra2019} (cyan circle). }\label{fig:shapes}
\vspace{-15pt}
\end{figure}

To quantitatively constrain the intrinsic 3D shapes, we consider triaxial models with axial lengths of $A>B>C$, parameterized by two axis ratios,  ellipticity ($\epsilon = 1-B/A$) and thickness ($C/A$; see Methods). The parameters that best reproduce the observations are derived by an Approximate Bayesian Computation MCMC through a MAP (Maximum A Posteriori) estimate (Methods), see 
red curves in Fig.~\ref{fig:hist-axis} (left column) and Extended Data Table~\ref{tab:properties}.
The full sample has negligible ellipticity $\langle B/A \rangle=0.87\pm0.06$, indicative of oblate structures, and thickness $\langle C/A \rangle=0.53\pm0.03$,
much higher than that ($\sim$0.1) of the submm and/or cold gas emission in local disks~\cite{Patra2019}, and higher than that ($\sim0.21\pm 0.02$) of their optical structures~\cite{Padilla2008}. The thickness is close to that ($\sim 0.4-0.5$) of Sloan high-luminosity, massive early-type spirals~\cite{Rodriguez2013}, which are known to be bulge-dominated. Placing these results in a schematic shape diagram taken from ref.~\cite{vanderwel2014b} (Fig.~\ref{fig:shapes}), the average submm-galaxy is indeed classified as a spheroid, fully consistent with Sloan early-type galaxies~\cite{Rodriguez2013}.

There is a significant dependence of triaxiality on compactness. The $\Sigma_{\rm SFR}$-compact subsample has significantly higher $\langle C/A \rangle=0.61\pm0.04$ than the $\Sigma_{\rm SFR}$-extended subsample ($\langle C/A \rangle=0.43\pm0.05$; Fig.~\ref{fig:hist-axis}).
These results show that most submm-bright galaxies  have round structures with a larger scale-height than those of present-day disk galaxies. The roundness appears to correlate with, and be driven by, their spatial compactness. Both the spheroidal shape and the compact, concentrated nature of the dense dust core in compact submm-galaxies suggest that we are observing bulge formation and, at the same time, assembly through in-situ dusty starbursts.

Our results differ from an earlier study~\cite{Gullberg2019} that, while finding evidence for triaxiality ($\langle C/A \rangle=0.28$), also recovered heavy ellipticity  ($\langle B/A \rangle=0.66$) and a median exponential profile ($n=1$),  concluding that submm-bright galaxies are disks with bars in the submm. We re-analysed those earlier, lower S/N ratio observations with our new technique, finding that for only 2/3rd of the galaxies, the morphology can be meaningfully modeled (the others fail, lacking sufficient S/N ratio). Their surface brightness profiles return a negative median Spergel-index, consistent with spheroids. Modeling their axis ratio distribution implies  $\langle B/A \rangle=0.85\pm0.08$ and  $\langle C/A \rangle=0.47\pm0.05$ (see Extended Data Fig.~\ref{fig:ext-gullberg}),  consistent with our higher S/N sample's results albeit with larger uncertainties. The different conclusions are thus ascribed to the imaging-plane-based analysis technique and to the limited S/N of the previous study (Methods).

Our selection criterion is independent of galaxy submm morphology, thus introducing no meaningful selection bias.
Only 2\% of the submillimeter-selected galaxies in our sample remain unresolved. 
It has been theorized that if submillimeter galaxies were disks, optical depth effects would render edge-on disks dimmer~\cite{Lovell2022}. This orientation bias would vary with wavelength, making elongated systems relatively less common at shorter wavelengths where optical depth is higher. We use observations at rest-frame wavelengths ranging from 123 to 630 $\mu$m, with a median of $236_{-38}^{+78}$ $\mu$m (uncertainties denote the interquartile range, see Extended Data Table~\ref{tab:properties}), where we primarily sample optically thin radiation in the Rayleigh-Jeans regime. Our sample does not exhibit a shortage of elongated submillimeter galaxies at shorter-wavelengths (see Extended Data Fig.\ref{fig:ext-lambda-snr}). Moreover, the slight increase in average axis ratio with S/N (see Extended Data Fig.\ref{fig:ext-lambda-snr}) stems from reduced uncertainties  (see Methods). Finally, any lensing effects from foreground objects or submillimeter emission originating from outer disks~\cite{LeBail2023} would diminish the observed axis ratios. Although these effects are challenging to quantify, accounting for them would reinforce our findings.

We investigate the cosmological relevance of submm-bright galaxies as a spheroid formation channel. The number density of  galaxies with SFR~$>620\ M_\odot$~yr$^{-1}$ at $z=2.7$ (both numbers are median values in our sample) is $4\times10^{-5}$~Mpc$^{-3}$~\cite{Traina2024}. Our sample spans $1.5$~Gyr of cosmic time (redshift semi-interquartile range of 2.0$-$3.4), corresponding to  2.7$\times10^{-5}$~Mpc$^{-3}$~Gyr$^{-1}$ forming events. The end-product of our sample is roughly at double its median stellar mass $M^*\sim10^{11}M_\odot$, assuming  50\% gas fractions in place at the moment of observing which imply gas consumption timescales of order of 100~Myr. We evaluate the formation rate of quiescent galaxies above $M^*\sim2\times10^{11}M_\odot$ by differentiating the evolving mass function\cite{Weaver2023}. This also requires $\approx2\times10^{-5}$~Mpc$^{-3}$~Gyr$^{-1}$ forming events, stable over $1.5<z<3.5$, suggesting that spheroid-forming submm-bright starbursts could be a dominant formation channel at cosmic noon. 

\begin{figure*}[htbp]
\centering
\includegraphics[width=0.3\linewidth]{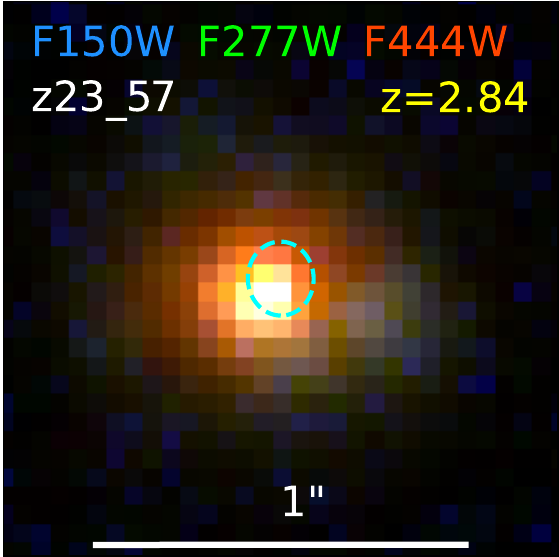}
\includegraphics[width=0.3\linewidth]{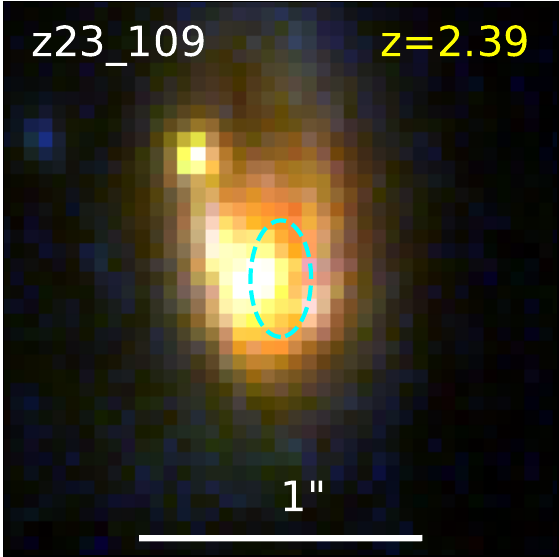}
\includegraphics[width=0.3\linewidth]{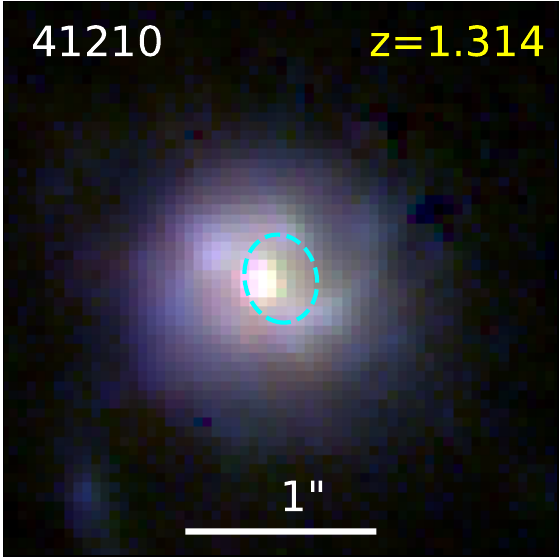}\\
\includegraphics[width=0.3\linewidth]{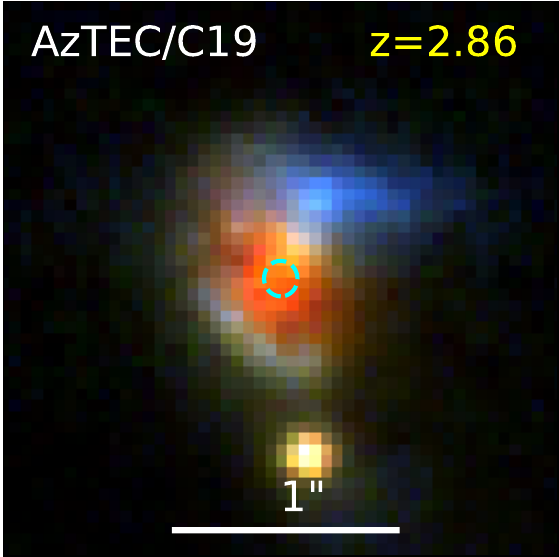}
\includegraphics[width=0.3\linewidth]{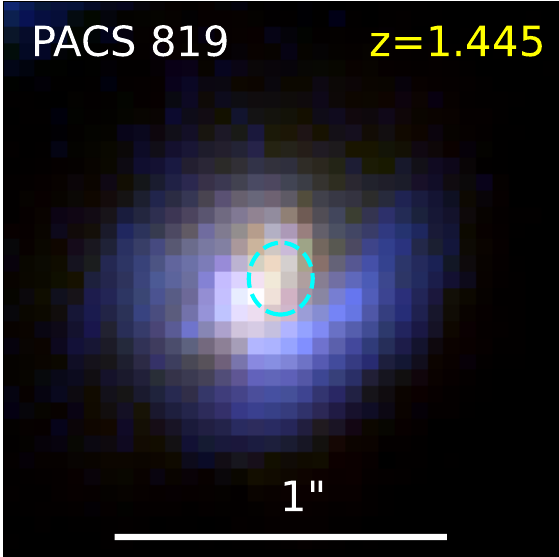}
\includegraphics[width=0.3\linewidth]{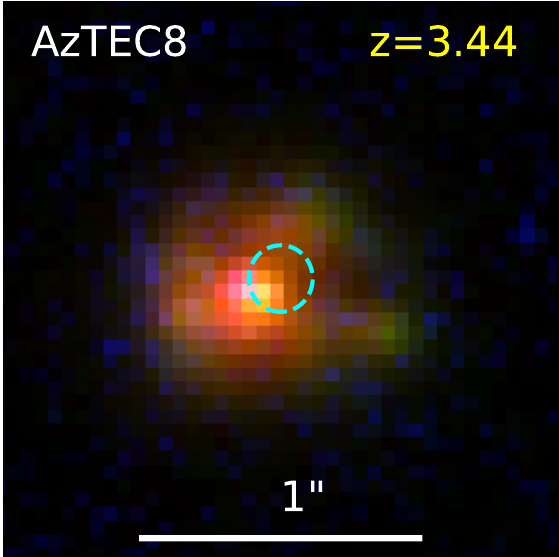}\\
\includegraphics[width=0.3\linewidth]{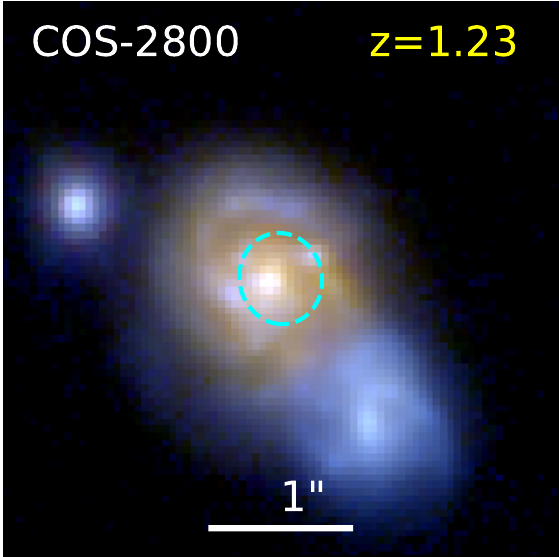}
\includegraphics[width=0.3\linewidth]{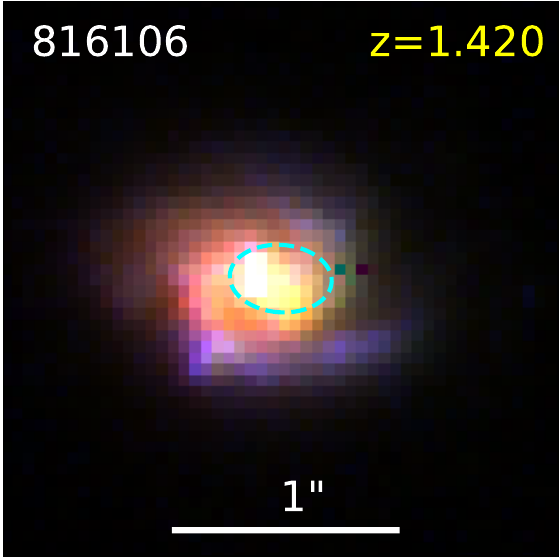}
\includegraphics[width=0.3\linewidth]{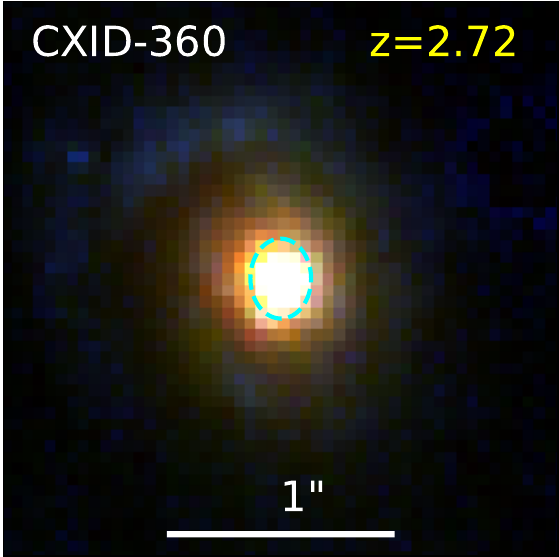}\\
\includegraphics[width=0.3\linewidth]{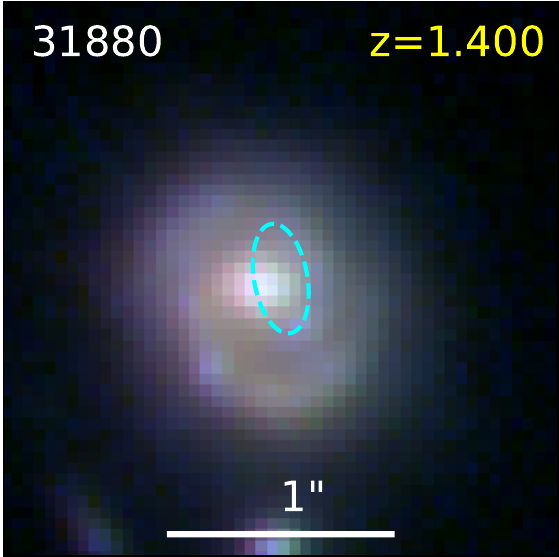}
\includegraphics[width=0.3\linewidth]{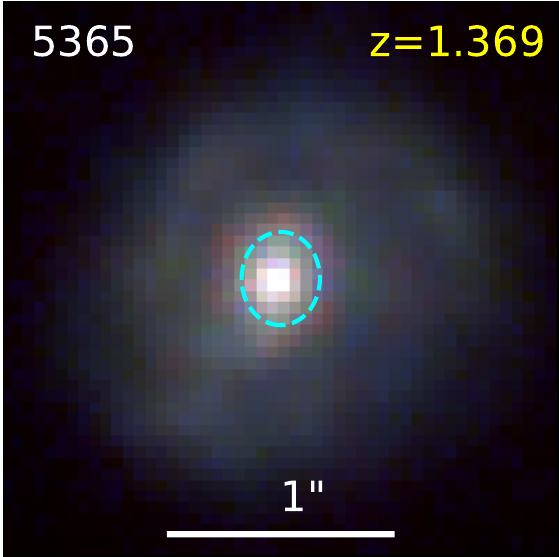}
\includegraphics[width=0.3\linewidth]{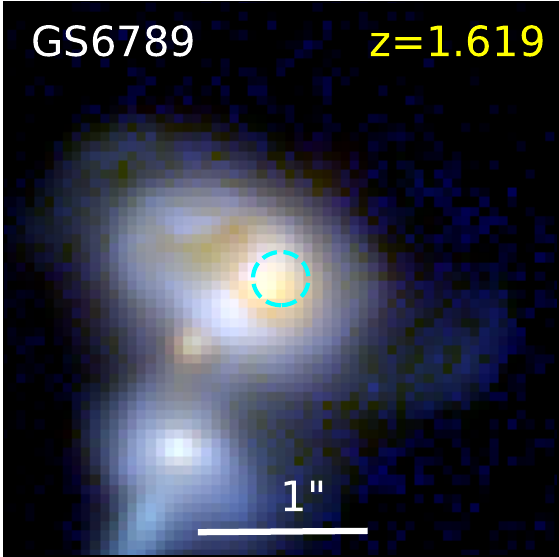}
\caption{\textbf{Examples of JWST cutout images for our sample of galaxies.} The RGB images are created using F444W for red, F227W for green, and F150W for blue. In each image, the cyan dashed region represents the best-fit for the submm emission using Spergel modeling, centered on submm emission's center. The elliptical shapes are plotted based on the submm $R_{\rm e}$, axis ratio, and the position angle derived from best-fitting by Spergel. The white bar indicates the scale of the image. The source name and redshift of the galaxies are indicated in each panel.}\label{fig:ext-jwst}
\end{figure*}

The tri-axial nature of the burst in submm-bright galaxies has further implications for the burst origin and spheroid formation. 
Disc instabilities, followed by the migration of clumps towards the nucleus\cite{Elmegreen2009, Dekel2009, Ceverino2015} primarily occur within the disk plane and may struggle to account for the spheroidal starburst. Conversely, counter-rotating spins resulting from tidal interactions and mergers~\cite{Toomre1977}, or non-co-planar accretion streams~\cite{Scannapieco2009, Sales2012, Aumer2013, Dubois2014}, can reduce the gas angular momentum producing compact spheroidal bursts. Simulations of major mergers incorporating high fractions of cold turbulent gas~\cite{Bournaud2011}, as opposed to those involving warm, stabilized gas, have  the ability to produce compact and round star-forming bursts (with C/A ratios exceeding 0.7-0.8), which align with our observations (see Extended Data Fig.~\ref{fig:hydro-simulations}). In these simulations, the kinematics can be dominated by dispersion, even within the gas component.
However, major mergers at coalescence are likely to disrupt pre-existing disk components~\cite{Bournaud2011}. Observations from the James Webb Space Telescope (JWST) show that massive disks always accompany~\cite{Kalita2022, LeBail2023,Gillman2023,Cardona-Torres2023} the compact starburst cores (heavily obscured by dust at optical wavelengths) in highly star-forming galaxies at $1.5<z<3$ (see Fig.~\ref{fig:ext-jwst}).
Misaligned cold streams, on the other hand, which are expected in massive galaxies~\cite{Kraljic2019}, would not disrupt the massive stellar disk components. Tidal perturbations from interactions or fly-bys could still be systematically required to enhance accretion, triggering starbursts with an expected duration of $\sim50$--100~Myr~\cite{Puglisi2019,Silverman2018b}. Steady accretion by itself would instead arguably persist over cosmological timescales.

We arrive at the same conclusion analysing EAGLE simulations~\cite{Schaye2015}. Following ref.~\cite{McAlpine2019}, we selected the most submm-bright galaxies at $z>1$, which return number densities, stellar masses, and offsets from the star-forming Main Sequence in agreement with our sample (Methods). The $z=0$ descendants of these galaxies are bulge-dominated galaxies with $[\alpha/{\rm Fe}]$ enhancement (Extended Data Fig.~\ref{Eagle_data}). At $z>1$, we find that only 10--50\% of the submm-bright galaxies in EAGLE had a recent merger, where the range of the fractions depends on whether the merger was in the last 100~Myr and was major (stellar mass ratio~$>1/4$) (i.e., could lead to a burst of star-formation), or if we allow for longer 600~Myr timescales and also count minor mergers (see Extended Data Fig.~\ref{Eagle_data}; Methods). This suggests that, based on theory, many submm-bright events are not merger-driven, in agreement with ref.~\cite{McAlpine2019}. On the other hand,  observations of widespread stellar mass asymmetries in distant starbursts~\cite{Cibinel2019,Kalita2022} point to the opposite conclusion. The contradiction might be solved, again, by recurring to tidal interactions being involved, since they could produce the observed mass asymmetries without being post-coalescence mergers. Quite importantly, the EAGLE submm-bright galaxies display highly increased accretion rates during 600~Myr preceding the bursts, with respect to control samples with the same stellar mass and redshifts (Extended Data Fig.~\ref{Eagle_data}), strongly supporting a primary role of gas accretion for in-situ bulge formation through starbursts. 

Stellar winds caused by these starbursts will likely prevent further substantial star formation and might induce quenching through feedback, at least in the core. Hence, for spheroids forming directly inside SMGs, the quenching of star formation will follow their assembly.  The already aging outer disks might finally contribute to the extended halos of the high \sersic \ profiles in elliptical galaxies if accretion stops~\cite{Setton2024}. Alternatively, early-type spirals might form in the disks that continue to be fed by incoming cold gas.

\begin{methods}

\section{ALMA galaxy sample}
We selected galaxies 
 from the A$^3$COSMOS database~\cite{Liu2019,Adscheid2024} (data version 20220606), including submm-bright galaxies observed by our ALMA projects in Band 6 (211--275 GHz) and 7 (275--373 GHz), with redshifts ranging from 0.9 to 5.9~\cite{Elbaz2018, Silverman2018, Rujopakarn2019, Valentino2020, Kalita2022, Jin2022}. To ensure accurate measurements of galaxy structure parameters (i.e., sizes, axis ratios, and \sersic \ indices), we only included galaxies with a detection significance of S/N$_{\rm beam}$ ($\equiv S_{\rm total}/\sigma_{\rm beam}$, the integrated flux density divided by the rms noise in the data) greater than 50. This high S/N threshold is required by our simulations designed to assess the robustness of the \sersic \ index measurements~\cite{Tan2024}. 
 
 We have identified 134 submm-bright galaxies in the COSMOS and GOODS-South fields that meet the criterion of having S/N$_{\rm beam}>50$.  Including the 12 galaxies with S/N$_{\rm beam}>50$ from ref.~\cite{Gullberg2019}, the full sample consists of 146 galaxies. Additionally, six galaxy systems with S/N$_{\rm beam}>50$ were identified as pairs of interacting or merging galaxies. In these cases, the S/N$_{\rm beam}$ measured for each individual galaxy is less than 50, so they were excluded from the analysis.

\section{Star formation rates, stellar masses, and redshifts}

We cross-matched the S/N-limited A$^3$COSMOS submm-bright galaxy catalog with the COSMOS2020 catalog~\cite{Weaver2022} and the COSMOS super-deblended catalog~\cite{Jin2018} using a search radius of 1.0$''$. This resulted in 72 counterparts in the optical-NIR and 80 in the far-infrared (FIR). All optical-NIR counterparts were identified in the FIR emission. We take stellar mass values from the COSMOS2020 catalog, estimated using classic aperture photometric methods, while SFR values are derived from the super-deblended catalog. The stellar masses and SFRs of galaxies in the GOODS-South field are obtained from the value-added galaxy catalog~\cite{Adscheid2024}. For the 24 galaxies selected from our published data~\cite{Elbaz2018, Silverman2018, Rujopakarn2019, Valentino2020, Kalita2022, Jin2022}, the stellar masses and SFRs are individually taken from the literature. All stellar mass and SFR values were rescaled to the Chabrier~\cite{Chabrier2003} initial mass function. 

In total, there are 39 galaxies in our sample with a spectroscopic redshift. For the rest of galaxies, we use the photometric redshifts from the COSMOS2020 catalog and value-added galaxy catalog. The median uncertainty of photometric redshift measured for our sample galaxies is 0.044, with a semi-interquartile range of 0.040. We utilize IR-to-radio photometric redshift derived from SED fitting~\cite{Jin2018} for the nine sources identified in FIR counterparts but without spectroscopic redshift and lacking photometric redshifts from COSMOS2020. For the other nine galaxies without counterparts in both optical and IR bands, we assume a redshift value of the median of 2.7. We find that excluding these nine sources from the statistical analysis would not significantly affect the conclusions of this work. The rest-frame wavelength of the sample galaxies ranges from 123 to 630~$\mu$m with a median value of $236_{-38}^{+78}$~$\mu$m. 

 The stellar mass and SFR of our sample galaxies with available measurements are in the range of $\sim 10^{10}-6.6\times 10^{11}\ M_\odot$ and $\sim 40-2800 \ M_\odot \ {\rm yr}^{-1}$, with a median of log($M_\star$)=$11.0_{-0.3}^{+0.2} \ M_\odot$ and SFR=$624_{-333}^{+429} \ M_\odot \ {\rm yr}^{-1}$, respectively. We calculate the specific SFR($\equiv$SFR/$M_\star$; sSFR) and plot the positions of the sample galaxies in the SFR$-M_\star$ MS plane (Extended Data Fig.~\ref{fig:ext-ms-plane}), which is the most typical representation of the correlation between SFR and $M_\star$ of star-forming galaxies~\cite{Daddi2007, Elbaz2007, Noeske2007}. We measured a median SFR-excess from the MS plane, $\Delta$MS (=log(SFR/SFR$_{\rm MS}$)), of $3.6_{-1.3}^{+3.5}$. We define starbursts as sources with a distance from the MS (which has a 1$\sigma$ scatter of $\sim$0.3 dex ) of $\Delta$MS $\geqslant$0.6 dex~\cite{Rodighiero2011}. We have 57 galaxies ($\sim 40\%$) classified as starbursts, while the rest of galaxies either fall within the scatter of the MS but above its average plane or lie close to our adopted starburst limit. For galaxies without IR counterparts in the COSMOS super-deblended catalog, the SFR were derived from the ALMA submm flux by assuming the median submm-flux to SFR ratio measured in our galaxy sample.

\section{Submm structure parameter measurements from the $uv$-plane}

To constrain the rest-frame FIR structure parameters (half-light radii, projected axis ratios, and \sersic \ indices) of our sample galaxies, we performed a visibility-based morphological analysis to model the galaxy light in dust continuum emission. The ALMA calibrated visibilities, obtained from A$^3$COSMOS pipeline using scripts provided by ALMA~\cite{Liu2019,Adscheid2024} and from our published data~\cite{Elbaz2018, Silverman2018, Rujopakarn2019, Valentino2020, Kalita2022, Jin2022}, were exported from CASA~\cite{McMullin2007} and imported to GILDAS~\cite{Guilloteau2000} using the \texttt{exportuvfits} and \texttt{fits$\_$to$\_$uvt} tasks, respectively. The four spectral windows were then combined using the \texttt{uv$\_$continuum} and \texttt{uv$\_$merge} tasks. The beam size, $\theta_{\rm b}$, for the observations was determined using natural weighting, which assigns each visibility a weight inversely proportional to the noise variance, maximizing point source sensitivity. We define $\theta_{\rm b}$ as the synthesized circularized beam size given by $\sqrt{ab}$, where {\it a} and {\it b} represent the full width at half maximums (FWHMs) of the major and minor axes of the synthesized beam, respectively.

Following the method described in detail by ref.~\cite{Tan2024}, we used the task \texttt{uv$\_$fit} to fit an elliptical Spergel model to the visibilities, allowing all fitted parameters (i.e., centroid position, flux density, effective radius, minor-to-major axis ratio, position angle, and Spergel index) to be free. The Spergel profile~\cite{Spergel2010} is an analytic alternative to the \sersic \ profile for modeling galaxy surface brightness profiles, which can be written as
\begin{equation}
\Sigma_\nu (R) = \frac{c_\nu^2 L_{\rm 0}}{R_{\rm e}^2} f_\nu \left( \frac{c_\nu R}{R_{\rm e}} \right)\label{eq:spergel},
\end{equation}
where $L_0$ is the total luminosity, $f_\nu(x) = \left( \frac{x}{2} \right)^\nu \frac{K_\nu (x)}{\Gamma (\nu + 1)}$, $\Gamma$ is the gamma function, $K_\nu$ is a modified spherical Bessel function of the third kind, $c_\nu$ is a constant, $R_{\rm e}$ is the half-light radius, and $\nu$ is known as the Spergel index that controls the relative peakiness of the core and the relative prominence of the wings (similar to  \sersic \ index  $n$), with a theoretical limit of $\nu > -1$. The Spergel profile has the advantage of having an analytic expression in both real space and Fourier space, making it easier to work with and to analyze in the \textit{uv}-plane.

Out of the 146 galaxies significantly detected in submm emission with S/N$_{\rm beam}>50$, 143 have robust constraints on their shape, with a median $R_{\rm e}/\Delta R_{\rm e}$ ratio of 10 and a median axis ratio \textit{q}($\equiv b/a$) of $q/\Delta q \sim$6. These 143 galaxies constitute the final sample of galaxies for the remainder of our analysis (See Extended Data Table~\ref{tab:measurements}), while the three unresolved galaxies ($\sim$2\%) were excluded from the analysis. For about 10\% of the galaxies that we cannot accurately measure the Spergel $\nu$ but can still measure $q$, we set the Spergel $\nu$ to the median value of the sample and incorporate errors from plausible variations of $\nu$. 

Extended Data Fig.~\ref{fig:ext-images} shows the dirty images and model-fitting images of our sample galaxies. A single Spergel profile can accurately represent the vast majority of our sample sources, while for sources that have an extended structure and/or multiple companion components, a combination of Spergel and point source or a combination of Spergel and Gaussian source models were applied. The uniform distribution of resulting residuals suggests that the source is well-described by the models. In Fig.~\ref{fig:ext-jwst}, we show JWST RGB cutout images of our sample of galaxies using the F150W, F277W, and F444W filters. We find that the submm emission tends to be predominantly concentrated in the central region of near-IR emission.

In the right panel of Extended Data Fig.~\ref{fig:ext-images}, we show the visibility profiles of each ALMA observation and the best-fit to the visibilities with a Spergel component model. The amplitude is extracted from the real part of the complex visibility in bins of $uv$-distance, which corresponds inversely to the observed angular scale.  For comparison, we overlay the best-fitting model with an exponential (\sersic \ $n=1$) profile. Both the best-fit Spergel and $n=1$ models were created with an extremely high S/N to ensure a robust measurement using the \texttt{uv$\_$fit} task. The non-flat visibility amplitude, showing a decreasing profile at long baselines, indicates that these sources are well-resolved. Based on the reduced chi-squared ($\chi^2_\nu$) statistic, the Spergel model provides a reasonably good representation of the submm emission, with a $\chi^2_\nu$ value close to 1, and demonstrates a superior fit compared to an exponential profile with lower $\chi^2_\nu$ for sources, particularly those with Spergel $\nu<0$. In the case of the galaxies shown as examples in Extended Data Fig.~\ref{fig:ext-images}, the $\chi^2$ distribution indicates that for the three galaxies with best-fit values of $\nu=-0.47$, $-0.6$, and $-0.73$, the probability ($P(n=1)$) corresponding to a degree of confidence that the \sersic \ $n=1$ solution can be rejected exceeds 95\%. This suggests a low likelihood that the submm emission in these galaxies conforms to an exponential disk distribution. For the other two galaxies with best-fit $\nu$ values of $-0.09$ and 0.65, the associated confidence level is 77.279\% and $<0.001\%$, respectively. These results imply a high probability that the dust emission in galaxies with $\nu>0$ follows a distribution similar to an exponential disk.

The flux density S/N$_{\rm beam}$ of our sample galaxies are in the range of $\sim 50-300$ with a median of 68$\pm$4. The median $\theta_{\rm b}$ is $0.51_{-0.31}^{+0.39}$~arcsec (uncertainties are interquartile range), and the measured median angular size $R_{\rm e}$ is $0.18_{-0.06}^{+0.07}$~arcsec. The size relative to the beam, $R_{\rm e}/\theta_{\rm b}$, ranges from 0.1 to 2.5, with a median of 0.41$\pm$0.04. The majority ($\sim 88\%$) of our sample of galaxies were observed in ALMA band 7 (median frequency of 344 GHz) with a median flux density of $7.8_{-4.7}^{+3.6}$~mJy (uncertainties are interquartile range), while for the remaining galaxies observed in ALMA band 6 (median frequency of 235 GHz), we measured a median flux density of $3.4_{-4.5}^{+0.5}$~mJy. The high submm flux measured indicates that most of our sample galaxies are extremely bright in submm emission, forming a fairly uniform SMG sample. 

For the model fitting of the visibilities, a reliable size measurement limit, $\theta_{\rm min}$, can be defined depending on the synthesized beam size $\theta_{\rm b}$ and the source S/N. This can be approximated as~\cite{Marti-Vidal2012}:
\begin{equation}
    \theta_{\rm min} = \beta \left( \frac{\lambda_{\rm c}}{2 (S/N)^2} \right)^{1/4} \times \theta_{\rm b} \simeq 0.88 \frac{\theta_{\rm b}}{\sqrt{S/N}},
    \label{eq:size-limit}
\end{equation}
with $\lambda_{\rm c}=3.84$ and $\beta$=0.75 as in ref.~\cite{Franco2018}. From our ALMA observations, the median $\theta_{\rm min}$ is 0.050$\pm$0.003~arcsec, with an interquartile range of 0.020$-$0.098~arcsec. For the whole SMG sample, we measured a median ratio of $R_{\rm e}$ to $\theta_{\rm min}$ of $3.9_{-1.6}^{+4.0}$, with uncertainties representing the interquartile range. Six sources have a measured size below the formal thresholds ($\theta_{\rm min}$), with a median $R_{\rm e}/\theta_{\rm min}$ of 0.8, based on Eq.~(\ref{eq:size-limit}). This indicates that the size measurements for the vast majority ($\sim 96$\%) of our sample galaxies are reliable also based on this metric. The small difference for the six galaxies is likely because our reliability assessment was not based solely on a generic relation with source S/N, but was derived from simulations~\cite{Tan2024} specific to our data and taking into account additional parameters of the galaxies. The reliability of the size measurement for these six galaxies is ensured by our simulations, validating their inclusion in the analysis. We also find that excluding these six sources from the statistical analysis would have no impact on the conclusions of this work.

The \sersic \ indices were accurately derived with an effective S/N of $n/\sigma_n > 3$. We convert the measured Spergel index $\nu$ into \sersic \ index \textit{n} using a relation inferred from simulations~\cite{Tan2024}:
\begin{equation}
    n(\frac{R_{\rm e}}{\theta_{\rm b}}, \nu) = 0.0249 \frac{R_{\rm e}}{\theta_{\rm b}} {\rm exp}(-7.72 \nu) + 0.191\nu^2 - 0.721 \nu + 1.32 \label{eq:conversion}
\end{equation}
where $R_{\rm e}$ is the effective radius and $\theta_{\rm b}$ is the circularized beam size. The Spergel index $\nu$ derived for the galaxies with robust measurements ranges from -0.86 to 1. We set an upper limit of \sersic \ index to $n=8$ for converting between \sersic \ $n$ and Spergel $\nu$. It should be noted that the conversion from Spergel $\nu$ to \sersic \ $n$ is not very sensitive to changes in $R_{\rm e}/\theta_{\rm b}$ at $\nu \geqslant 0$. For example, for galaxies with a Spergel index of $\nu=0$, the corresponding \sersic \ $n$ values, converted based on Eq.~(\ref{eq:conversion}), for both compact-sized ($R_{\rm e}/\theta_{\rm b}=0.1$) sources and significantly more extended ($R_{\rm e}/\theta_{\rm b}=2$) sources, are comparable with a value of $n\sim 1.3-1.4$.

Employing a standard $\Lambda$CDM cosmology ($H_0=70$~km~s$^{-1}$~Mpc$^{-1}$, $\Omega_{\rm m}=0.3$, $\Omega_\Lambda=0.7$), we find the submm sizes of the sample galaxies ranging from 0.4 to 5.4 kpc with a median of $R_{\rm e}=1.4_{-0.5}^{+0.6}$ kpc. Among these galaxies, a subset of 71 galaxies showing SFR surface density higher than the average value at their redshifts, are found to be compact in submm emission with a median $R_{\rm e}$ of $0.9_{-0.2}^{+0.4}$ kpc. We define this subsample of galaxies with $\Sigma_{\rm SFR}$ higher than the average trend as $\Sigma_{\rm SFR}$-compact galaxies in this study. For the rest of galaxies in our sample, the submm emission is more extended with a median $R_{\rm e}$ of $1.8_{-0.4}^{+0.8}$ kpc, which we refer to as $\Sigma_{\rm SFR}$-extended galaxies. In addition, we measure a median $R_{\rm e}$ of $1.5_{-0.5}^{+0.7}$ kpc for the pure disks with Spergel index $\nu>0$, which is slightly larger than the median of $1.3_{-0.4}^{+0.5}$ kpc measured for the galaxies with $\nu<0$. With the measured submm $R_{\rm e}$, we find the SFR surface density $\Sigma_{\rm SFR}$, calculated as 0.5$\times$SFR/($\pi R^2_{\rm e}$), of our sample are in an interquartile range of $\sim 19 - 122$~$M_\odot$~yr$^{-1}$~kpc$^{-2}$ with a median value of $49_{-30}^{+73}$~$M_\odot$~yr$^{-1}$~kpc$^{-2}$. The median $\Sigma_{\rm SFR}$ measured for the subsample of compact and extended galaxies are $111_{-30}^{+110}$ $M_\odot$~yr$^{-1}$~kpc$^{-2}$ and $24_{-13}^{+15}$ $M_\odot$~yr$^{-1}$~kpc$^{-2}$, respectively. In addition, we found that the distribution of $\nu$ measured for the two subsamples are significantly different (see Extended Data Fig.~\ref{fig:ext-nu-distribution}). 

\section{Modeling the distribution of Spergel index $\nu$}\label{subsec: modeling-nu}

Figure~\ref{fig:histogram}\textbf{e} shows the distribution of Spergel index $\nu$ measured for the whole sample of galaxies. To account for the noise in the $\nu$ measurements, we model the observed distribution of $\nu$ to infer the intrinsic distribution by assuming a two-Gaussian model. We performed Monte Carlo (MC) simulations to obtain uncertainty estimates on $\nu$. Following the method described in ref.~\cite{Tan2024}, each simulated galaxy consists of a model source signal inserted in an empty dataset with realistic noise that is from our ALMA real data. In the simulation, we fix the flux density S/N, size $R_{\rm e}/\theta_{\rm b}$, and axis ratio to the median values measured for our ALMA sample galaxies, while we vary the Spergel $\nu$ from -0.8 to 0.7. 

Extended Data Fig.~\ref{fig:ext-sigma-function}\textbf{a} shows the measurement uncertainty of Spergel $\nu$ as a function of $\nu$. The measurement uncertainty is evaluated as the median absolute deviation (MAD) of the data around the true value converted to the standard deviation $\sigma$ using $\sigma$=1.48$\times$MAD, which is what is expected for a Normal distribution. The use of MAD is more robust to outliers, while capturing the bulk of the spread in the sample. By defining the deviation with respect to the true value we are taking the measurement bias into account as well to evaluate the overall uncertainty in the measurements. The simulation shows that the uncertainties in $\nu$ measurements are significantly higher at $\nu >0$ than at $\nu < 0$, with an average $\sigma_\nu$ of $\sim 0.8$ at $\nu=0.5$ and of $\sim 0.2$ at $\nu=-0.5$. The best linear least-squares fit to the simulation data yields
\begin{equation}
    \sigma_\nu = 0.60(\pm 0.05)\nu + 0.55(\pm0.04).
    \label{eq:sigma_nu}
\end{equation}

We performed an Approximate Bayesian Computation (ABC) analysis using the python Markov chain Monte Carlo (MCMC) package \texttt{emcee}~\cite{Foreman-Mackey2013} to search for best-fitting model, which minimize the differential between the CDF of the data and that of the model. To account for the uncertainty of $\nu$ measurement, we generate 10000 realizations of the data for each model $\nu$ by assuming a Gaussian distribution for $\sigma_\nu$. In MCMC, each time a model distribution of $\nu$ values is generated and convolved with a series of Gaussians with the function defined from simulation (Eq.~(\ref{eq:sigma_nu})). 

The best-fit $\nu$ distribution of the sample galaxies when accounting for the errors and the deconvolved distribution of $\nu$ are indicated by the orange and blue curves in Fig.~\ref{fig:histogram}\textbf{e}, respectively. The best-fit mean $\mu$ and standard deviation $\sigma$ for the two Gaussian components are $\mu_1 = -0.34_{-0.02}^{+0.04}$, $\sigma_1=0.03_{-0.02}^{+0.04}$, and $\mu_2 = 0.46_{-0.28}^{+0.19}$, $\sigma_2=0.02_{-0.01}^{+0.16}$, respectively. The best-fit $f_{\nu > 0}$ is $0.18_{-0.09}^{+0.08}$, representing the fraction of galaxies classified as pure galaxies with $\nu>0$ (see Extended Data Fig.~\ref{fig:ext-mcmc-corner}). For the subsamples of $\Sigma_{\rm SFR}$-compact and -extended galaxies, the best-fit $\nu$ distribution indicates a fraction of pure disks of $12_{-7}^{+13}$\% and $19_{-12}^{+24}$\%, respectively.

\section{Modeling the intrinsic 3D shape distribution}\label{subsec: modeling-q}

We model the observed, projected axis ratio distribution to infer the intrinsic 3D shape by assuming a triaxial ellipsoid model, which is characterized by major axis $A$, intermediate axis $B$, and minor axis $C$. For disk galaxies, the ratio of $B/A$ can be regarded as a measure for the ellipticity ($\epsilon \equiv 1-B/A$) and $C/A$ quantifies the relative system thickness encompassing all galactic components, including disks, bars, and bulges. We compute the apparent axis ratio \textit{q} at a random viewing angle ($\theta, \phi$) using Equations (11) and (12) in ref.~\cite{Binney1985}
\begin{equation}
    q = \sqrt{\frac{V + Z - \sqrt{(V - Z)^2 + W^2}}{V + Z + \sqrt{(V - Z)^2 + W^2}}},
    \label{eq:q}
\end{equation}
where
\begin{equation}
    V = \frac{{\rm cos}^2\theta}{\gamma^2} \left({\rm sin}^2\phi + \frac{{\rm cos}^2\phi}{\beta^2}  \right) + \frac{{\rm sin}^2\theta}{\beta^2}, \nonumber
\end{equation}
\begin{equation}
    W = {\rm cos}\theta {\rm sin}2\phi \left(1 - \frac{1}{\beta^2}\right)\frac{1}{\gamma^2}, \nonumber
\end{equation}
\begin{equation}
    Z = \left(\frac{{\rm sin}^2\phi}{\beta^2} + {\rm cos}^2\phi  \right)\frac{1}{\gamma^2}, \nonumber
\end{equation}
and $\beta = B/A$ and $\gamma = C/A$. The viewing angle ($\theta, \phi$) is in the spherical polar coordinate system, where the random viewing positions are distributed uniformly in the cosine of the polar coordinate $-90^\circ \leqslant \theta \leqslant 90^\circ$ and  the azimuthal coordinate $0^\circ \leqslant \phi < 360^\circ$.

To account for variations in intrinsic 3D shape, we assume a lognormal distribution for the face-on ellipticity $\epsilon$ with a mean of $\mu (={\rm ln} \epsilon)$ and dispersion of $\sigma$, and a Gaussian distribution for the edge-on thickness $C/A$ with mean $\mu_\gamma$ and standard deviation $\sigma_\gamma$, respectively, following ref.~\cite{ryden2004}. For a given set of parameters ($\mu$, $\sigma$, $\mu_\gamma$, $\sigma_\gamma$), we compute the apparent axis ratio $q$ at a random viewing angle ($\theta, \phi$). The distributions of model axis ratios were calculated in a range of $\mu = -5.00 \sim -0.05$, $\sigma = 0.20\sim 4.00$, $\mu_\gamma = 0.10\sim 1.00$, and $\sigma_\gamma = 0.01\sim 1.00$. Repeating this calculation, we obtain a model distribution of apparent $q$, which is used to compare with the observed distribution of our sample.

To assess the reliability of our axis ratio \textit{q} measurements, we performed MC simulations to obtain uncertainty estimates on \textit{q}. Similar to the method used to estimate the uncertainties in the Spergel $\nu$ measurements, we fix the flux density S/N, size $R_{\rm e}/\theta_{\rm b}$, and Spergel index to the median values measured for our ALMA sample galaxies, but vary the axis ratio $q$ uniformly from 0 to 1 in the simulation. Extended Data Fig.~\ref{fig:ext-sigma-function}\textbf{b} shows the measurement uncertainty of the apparent axis ratio \textit{q} as a function of \textit{q}. The measurement uncertainty is evaluated as the MAD of the data around the true value converted to the standard deviation $\sigma$ using $\sigma$=1.48$\times$MAD. The simulation shows that the uncertainty of \textit{q} tends to be higher for values of \textit{q}$\leqslant$0.3 than those of \textit{q}$>$0.3, with $\sigma_q \sim 0.15$ at \textit{q}$\sim$0 and $\sigma_q \sim 0.10$ at \textit{q}$\sim$1. A linear fit to the simulation data gives a relation with 
\begin{equation}
    \sigma_q = -0.04(\pm 0.02)q + 0.13(\pm 0.01).
    \label{eq:sigma_q}
\end{equation}

We performed an ABC analysis to search for best-fitting models. To account for the uncertainty of \textit{q} measurement, we generate 10000 realizations of the data for each model \textit{q} by assuming a Gaussian distribution for $\sigma_q$. In MCMC, each time a model distribution of \textit{q} values is generated and convolved with a series of Gaussians with the function defined from simulation (Eq.~(\ref{eq:sigma_q})).

The red curves in the left column of Fig.~\ref{fig:hist-axis} represent the best-fit model with the highest posterior probability (akin to Maximum A Posteriori estimate), which is identified as the step with highest posterior probability and nearest to the position defined by averages $\langle B/A \rangle$ and $\langle C/A \rangle$ (the black squares in the right panels). This is referred to as the MAP best-fit model in our work.  The right panels of Fig.~\ref{fig:hist-axis} and Extended Data Fig.~\ref{fig:ext-mcmc-corner} show the posterior probability distribution of the fitting parameters from the MCMC analysis. The average best-fit values of $\langle B/A \rangle$ and $\langle C/A \rangle$ for the full sample are $0.87\pm0.06$ and $0.53\pm0.03$, respectively. The best-fit $\langle B/A \rangle$ and $\langle C/A \rangle$ for the subsamples of $\Sigma_{\rm SFR}$-compact and $\Sigma_{\rm SFR}$-extended galaxies are summarized in Extended Data Table~\ref{tab:properties}.

\section{Distribution of axis ratios across various compact sub-samples}

In addition to the subsample of galaxies classified based on the SFR surface density, we have also examined the modeling of axis ratios for subsamples split using different methods. These methods include defining a subsample of galaxies classified as submm compact galaxies, characterized by $R_{\rm e}<1$~kpc or $R_{\rm e}$ smaller than the average trend ($\langle R_{\rm e}(z)\rangle$) at their respective redshifts to establish the boundary. Similar to the results obtained for the subsample of $\Sigma_{\rm SFR}$-compact galaxies, the best-fitting model shows a significantly higher disk thickness, with $\langle C/A \rangle \sim 0.6$ for the sample of compact galaxies selected based on either  $R_{\rm e}<1$~kpc or $R_{\rm e}<\langle R_{\rm e}(z)\rangle$, indicating an intrinsic shape closer to spheroidal. For the remaining galaxies with a more extended structure in submm emission, the MAP best-fit $\langle C/A \rangle$ is $0.4-0.5$. It is important to note that for the sample of $\Sigma_{\rm SFR}$-compact galaxies, the MAP best-fit model also has a lower $\chi^2$ value compared to other methods used to define submm compact galaxies, and thus it was chosen as the primary one discussed in the paper and figures. In addition, there are reasons to believe that $\Sigma_{\rm SFR}$ is indeed a key parameter regulating galaxy properties~\cite{Narayanan2014, Rujopakarn2011}.  

To test the heterogeneity in the depths and the range of beam sizes of the observations that might impact the results, we also split the sample at the median flux and the median beam size, respectively. Extended Data Fig.~\ref{fig:ext-subsamples} shows the observed $q$ distribution and the triaxial modeling results for both subsamples. The distributions and the measured $\langle B/A \rangle$ and $\langle C/A \rangle$ are consistent within the uncertainties for the subsamples split either by flux or beam size, suggesting that both depth effects and the diversity of observational setups are unlikely to affect our results substantially.

\section{Comparison with previous studies}

An earlier study~\cite{Gullberg2019} examined the distribution of axis ratio in a comparably sized sample of 153 S/N$>$8 submm galaxies, finding best-fit B/A and C/A distributions peaking at 0.68$\pm$0.02 and 0.28$\pm$0.01, respectively. We reanalyzed the ALMA dataset from ref.~\cite{Gullberg2019} using our $uv$-plane-based method. The S/N$_{\rm beam}$ estimates for the sample in ref.~\cite{Gullberg2019} range from 7.4 to 68.0 with a median of 30.4$\pm$0.9, with 12 galaxies meeting our selection criterion of S/N$_{\rm beam}>50$. The profile fitting to the sample using a Spergel model shows that only 99 ($\sim$65\%) galaxies can be resolved with the criterion of $R_{\rm e}/\Delta R_{\rm e} > 3$. Based on this subsample, we performed the corresponding statistical analysis. We find a median Spergel $\nu$ of -0.16$\pm$0.06, with $\sim 20$\% of the sample galaxies consistent with pure disks (Extended Data Fig.~\ref{fig:ext-gullberg}). The distributions of $\nu$ and $q$ are similar to those of our $\Sigma_{\rm SFR}$-extended subsample. When modeling the triaxiality of these galaxies, we inferred $\langle B/A \rangle$ of 0.85$\pm$0.08 and $\langle C/A \rangle$ of 0.47$\pm$0.05, respectively. Comparing the individual measurements for these galaxies, we find that the \sersic \ index measurements converted from Spergel $\nu$~\cite{Tan2024} are higher than those reported in ref.~\cite{Gullberg2019}, with a median value of 1.5$\pm$0.1. In addition, we have a larger proportion of galaxies fitted with $q>0.8$ compared to those measured by ref.~\cite{Gullberg2019}. Both of these biases are largely expected given the simulations from ref.~\cite{Tan2024} comparing $uv$- versus image-based measurements at the typical S/N of the sample in ref.~\cite{Gullberg2019}. Overall, all these biases, along with the fact that ref.~\cite{Gullberg2019} measured parameters for all 153 galaxies, including those with very low S/N, can explain the different conclusions we obtained from reanalysing the dataset and reinforce the findings of our work.

\section{Comparison with local (U)LIRGs}

To explore the possible processes that trigger the star formation activity in submm compact galaxies, we compare the dust continuum structures with those of (U)LIRGs in the local Universe, where the starburst is typically concentrated within the central $\sim 1$~kpc. Moreover, the star formation surface density $\Sigma_{\rm SFR}$ measured for the local ULIRGs and submm compact galaxies are found to be comparable ($>$ 100 $M_\odot\ {\rm yr}^{-1}\ {\rm kpc}^{-2}$) (e.g., ref.~\cite{Barcos-Munoz2017}; see Extended Data Table~\ref{tab:properties}). In Fig.~\ref{fig:histogram}\textbf{c} we show a comparison of axis ratio $q$ distributions between local (U)LIRGs~\cite{Ueda2014} and our sample of galaxies. Given that the inclination of (U)LIRGs in ref.~\cite{Ueda2014} is estimated by fitting the velocity field of CO molecular gas with tilted concentric ring models, we calculate the observed \textit{q} corresponds to the inclination for local (U)LIRGs as $q=$~cos~$i$ by assuming, conservatively, that the authors derived inclinations assuming very thin disks. A two-sample Kolmogorov-Smirnov test between the subsample of submm compact galaxies and local (U)LIRGs gives a $p-$value of 0.39, suggesting that we cannot reject the null hypothesis that these two samples are drawn from the same parent distribution. Larger samples of local ULIRGs are required to substantiate this suggestion. 

\section{Implications for kinematic studies}

Submm galaxies often show evidence of velocity gradients, suggesting overall rotation~\cite{Barro2017, Xiao2022, Lelli2023, Rizzo2023, Amvrosiadis2023, Liu2023, Birkin2023}. This is not inconsistent with their 3D shapes being spheroidal-like. 
Even in the local Universe important rotational support is often inferred in early-type galaxies~\cite{Cappellari2016}. 
Generally, kinematic studies assume that submm galaxies are intrinsically flat disks, leading to underestimates of their inclination and, thus, overestimates of rotational velocity. The results of this paper suggest that intrinsic rotational velocities might be lower than typically estimated. Deriving the actual inclination of spheroidal systems is difficult, but their 3D shapes could be further constrained by looking at the radial dependence of observed velocities. We caution though that comparably large and high S/N samples of CO emission line maps are not currently available. It has been shown that often galaxies are larger in CO than in continuum, providing possibly different views of the host galaxy. It would be important in the future to verify these results with CO.

\section{Analysis of the EAGLE simulations}

We use the EAGLE cosmological simulation (100 cMpc cosmological box; refs.~\cite{McAlpine2016,Schaye2015}) to investigate the merger history of submm galaxies. We create a sample of submm galaxies by selecting galaxies at $z\geq 1$ with mock submm fluxes of $S_{850\mu\rm{m}}{\geq}$ 1 mJy. These fluxes at 850 $\mu$m were computed in post-processing using the radiative transfer code (SKIRT), after assuming a dust-to-metals ratio~\cite{Camps2018,McAlpine2019}. Extended Data Fig.~\ref{Eagle_data}\textbf{a} shows the stellar mass as a function of redshift for all submm galaxies in the cosmological box, whereas the panel \textbf{b} highlights the offset in SFR of the submm sample relative to the star-forming Main Sequence of the simulation at $z=1$ (blue solid line). In the latter, the SFRs of individual submm galaxies where renormalized to match their expected values at $z=1$, revealing that the SFRs of the submm sample deviate by more than 2$\sigma$ from the median relation (as highlighted by the shaded region). The panel \textbf{c} shows the fraction of submm galaxies that experienced a major (stellar mass ratio $>1/4$, blue symbols) or major plus minor merger (stellar mass ratio $>1/10$, grey symbols) over the preceding 600 Myr (dotted symbols) or 100 Myr (crosses).  We find that between 30\% and 40\% of submm galaxies at $1<z<2$ have undergone a major merger when analysing their merger history across a span of 600 Myr. This fraction increases to 60\% and 100\% for $z>2$. Mergers seem to dominate for $z>3$ galaxies, but not necessarily for $z<3$ galaxies. When galaxies are only tracked for 100 Myrs, the merger fraction decreases. We find that only $10\%$ of submm galaxies at $z=2.5$ underwent a major merger during the previous 100 Myrs. This suggests that a significant fraction of submm-bright events are not driven by mergers. 

Panel \textbf{d} from Extended Data Fig.~\ref{Eagle_data} shows the rate of gas accretion approximately 600 Myr before galaxies became submm-bright. To calculate this rate, we compared the total gas mass within a 3D aperture of 50 kpc around the submm galaxies relative to their progenitors in the previous output redshift. In the panel, orange symbols highlight the accretion rates of submm galaxies, while blue symbols show the median (along with the 16/84th percentiles) gas accretion rates of a control sample. This control sample consists of galaxies with stellar masses within $\pm 0.2$ dex of the median stellar mass of the submm galaxies at each output redshift. The panel indicates that submm galaxies accreted up to ten times more gas mass per year than typical galaxies of the same mass, suggesting that the submm phase was likely triggered by a high rate of gas inflow.

Panels \textbf{e} and \textbf{f} explore the morphology and [$\alpha$/Fe] ratios of the galaxies that the submm sample evolves into. Panel \textbf{e} shows the Disc-to-Total (D/T) mass ratio of all galaxies at $z=0$ (blue symbols) and the submm descendants (orange symbols). Similarly, panel \textbf{f} shows the stellar [$\alpha$/Fe] ratio (represented by [O/Fe]). These panels indicate that submm galaxies evolve into elliptical galaxies with typical [$\alpha$/Fe] element ratios.

\end{methods}

\clearpage

\begin{extdata}

\setcounter{figure}{0}
\begin{figure*}[htbp]%
\centering
\includegraphics[width=\linewidth]{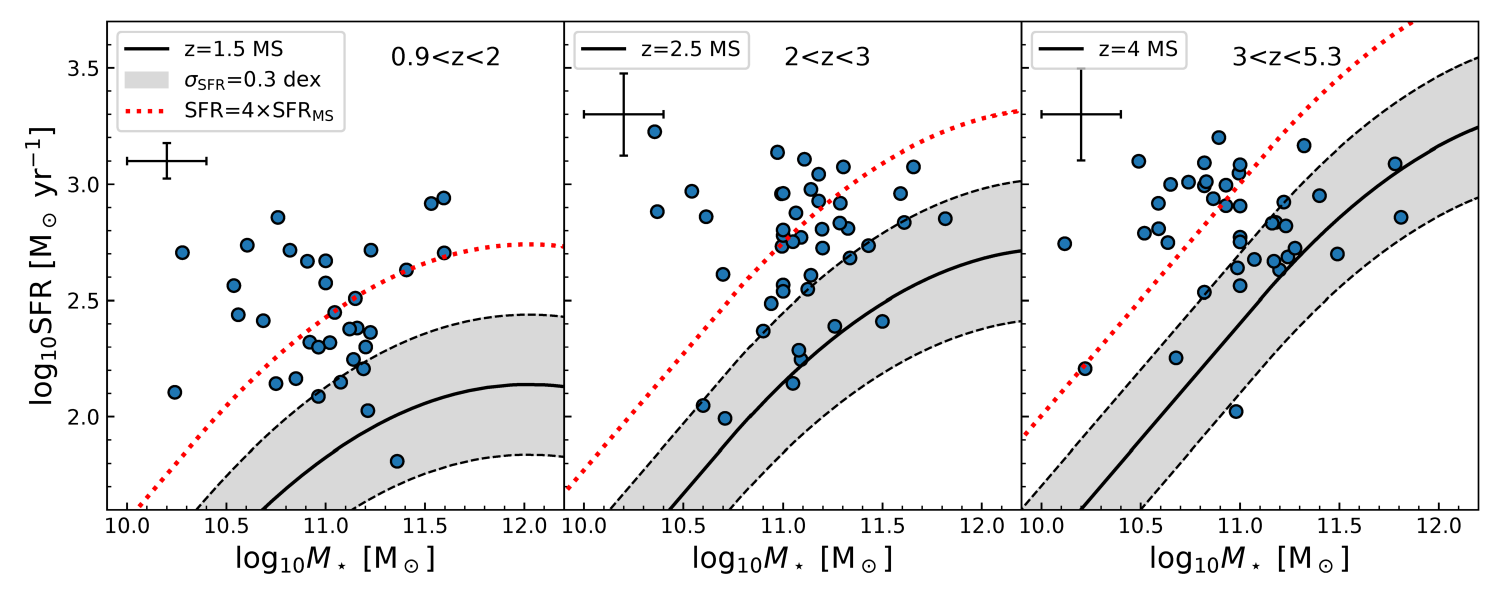}
\captionsetup{labelfont=bf,name=Extended Data Fig.,labelsep=period}
\caption{\textbf{Sample of ALMA submm-bright galaxies in the SFR-$M_\star$-$z$ plane.} The black solid and dashed lines represent the positions of main sequence at a common redshift (from left to right: $z=$1.5, 2.5, and 4) and the associated 1$\sigma$ dispersion given by ref.~\cite{Schreiber2015}. The red dotted line shows the MS threshold above which galaxies are classified as starbursts (SFR/SFR$_{\rm MS}>4$, $\Delta{\rm MS}$=0.6 dex). We note that the values of SFR shown in each panel were scaled to a common redshift $z_{\rm bin}$ by multiplying the actual SFR by a factor of SFR$_{\rm MS}^{z_{\rm bin}}$/SFR$_{\rm MS}^z$, in order to maintain the relative position of each galaxy with respect to the MS at its redshift. The vertical and horizontal bars indicate the median uncertainties of $M_\star$ and SFR, respectively.}
\label{fig:ext-ms-plane}
\end{figure*}

\begin{figure*}[htbp]%
\centering
\includegraphics[width=0.7\linewidth]{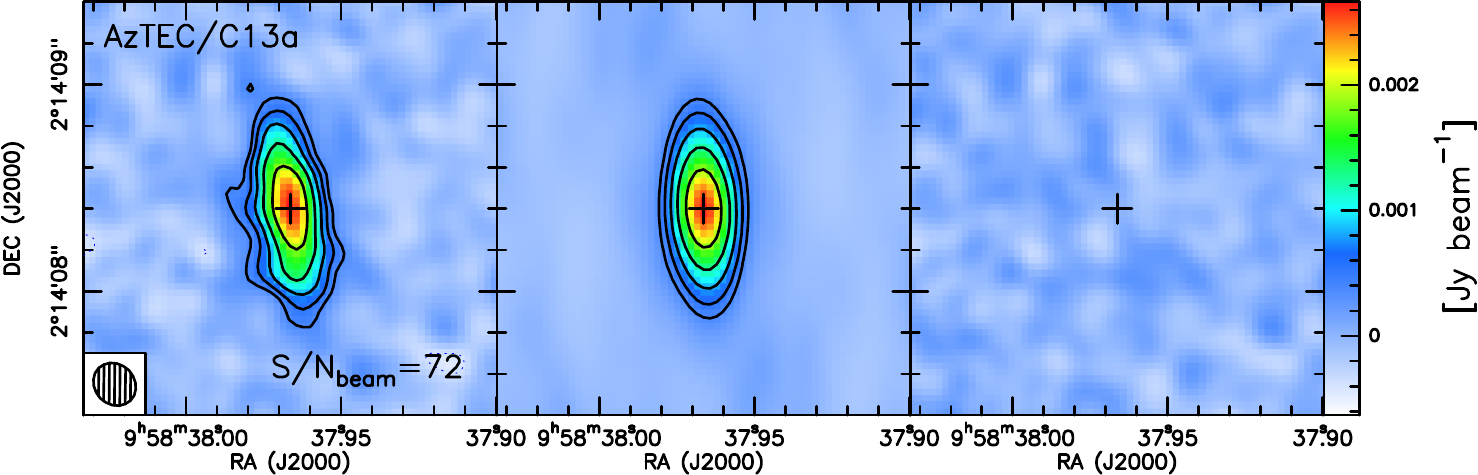}
\includegraphics[width=0.28\linewidth]{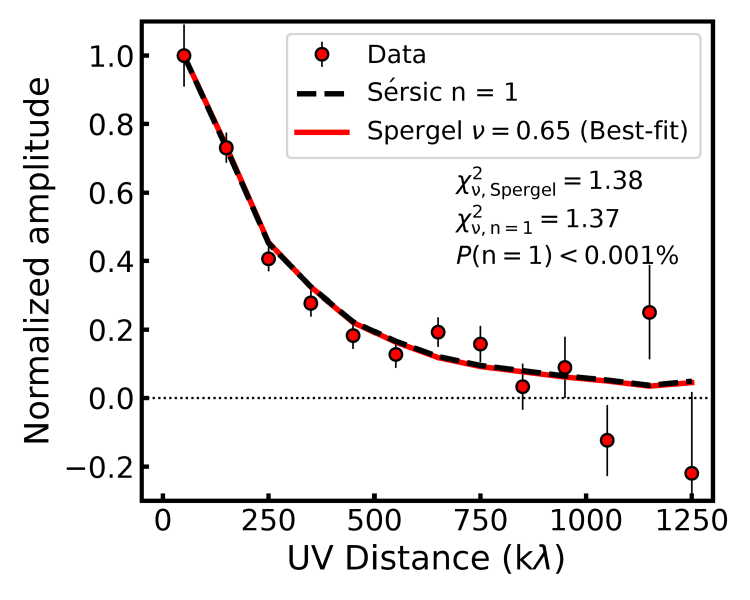}
\includegraphics[width=0.7\linewidth]{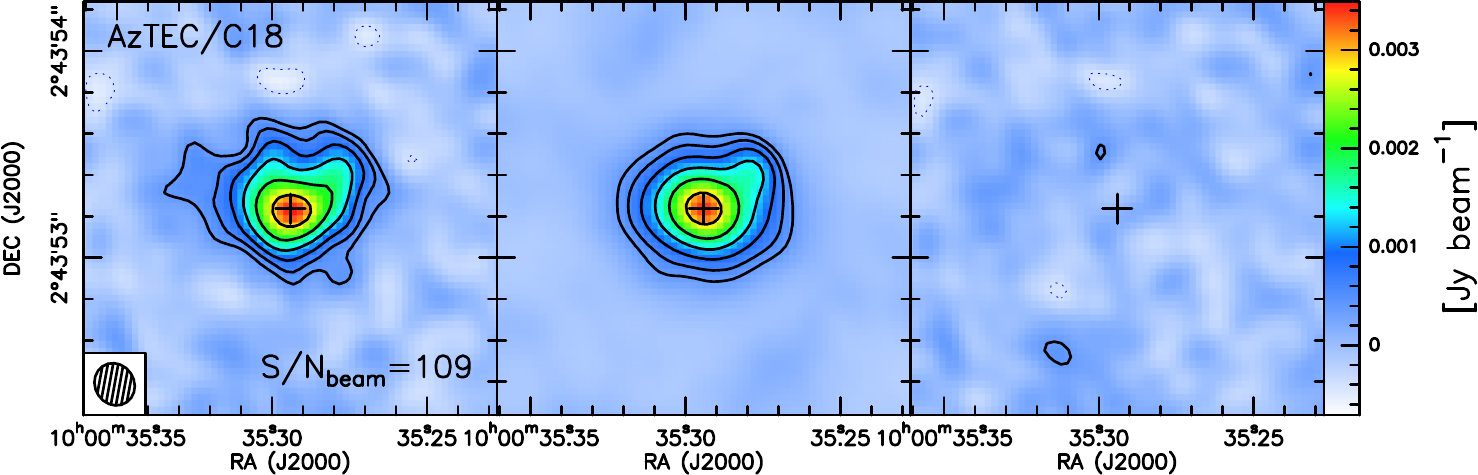}
\includegraphics[width=0.28\linewidth]{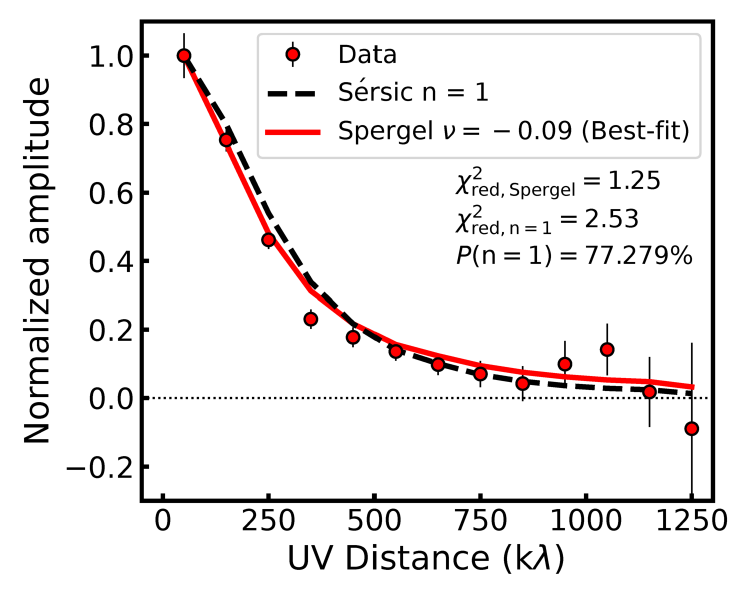}
\includegraphics[width=0.7\linewidth]{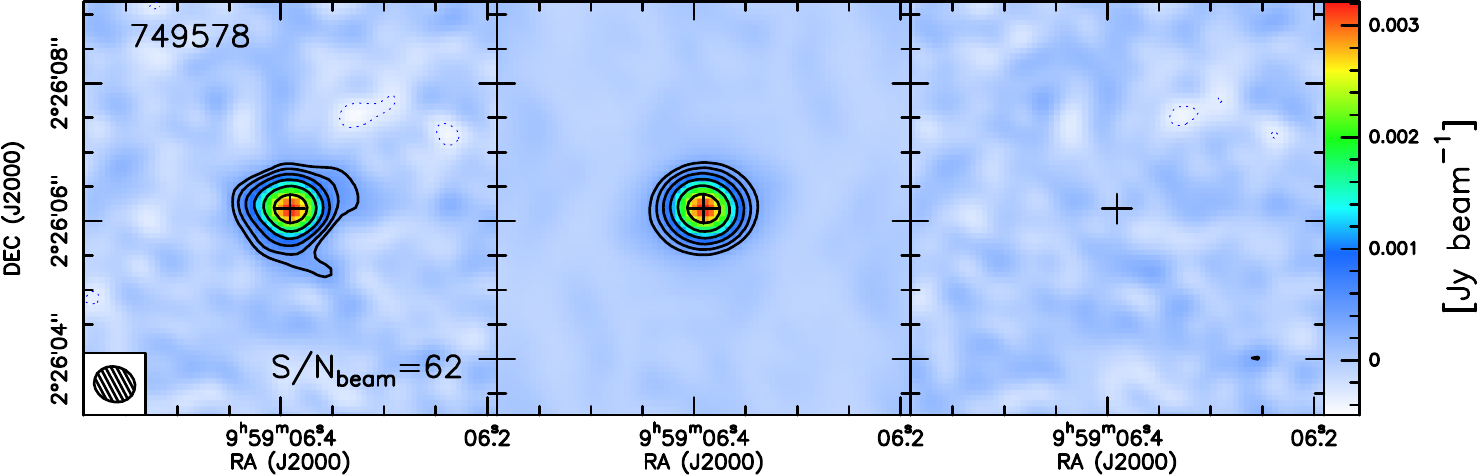}
\includegraphics[width=0.28\linewidth]{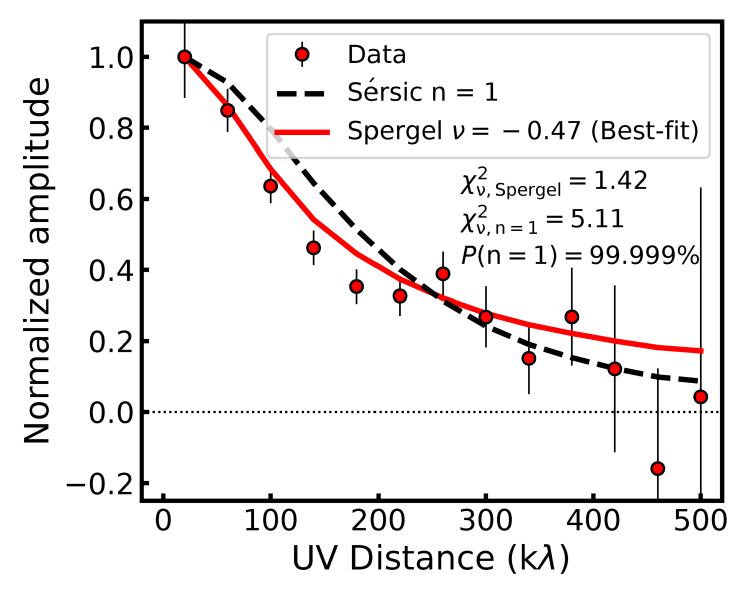}
\includegraphics[width=0.7\linewidth]{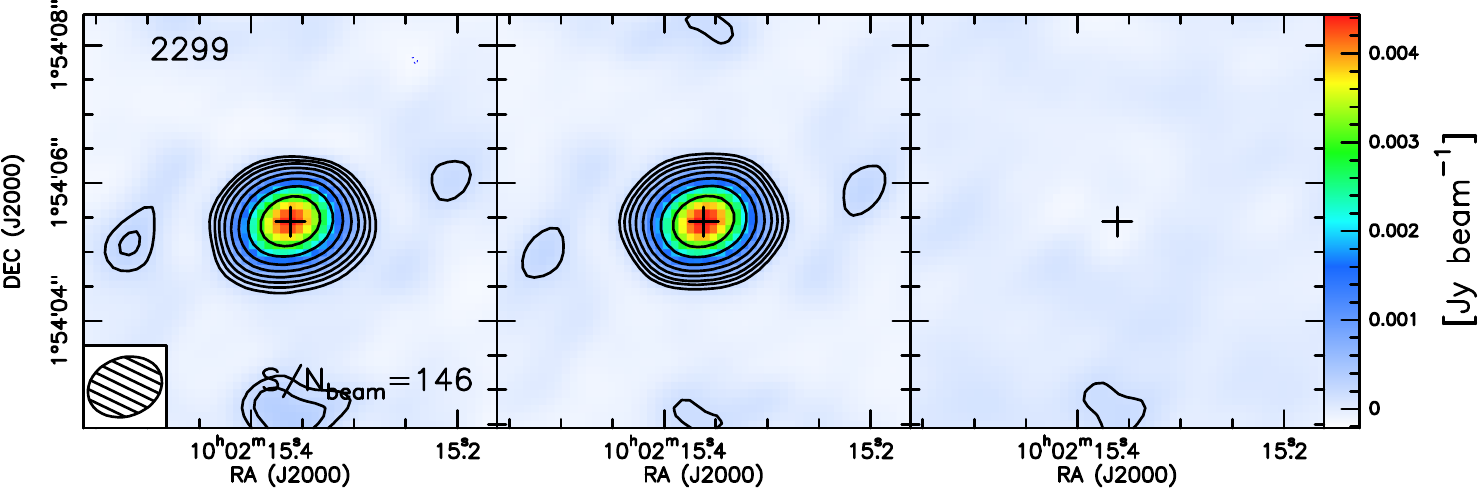}
\includegraphics[width=0.28\linewidth]{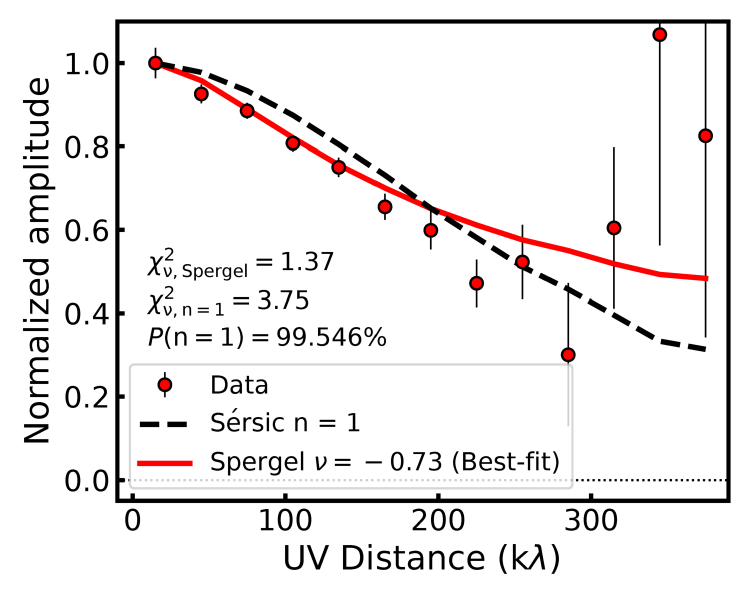}
\includegraphics[width=0.7\linewidth]{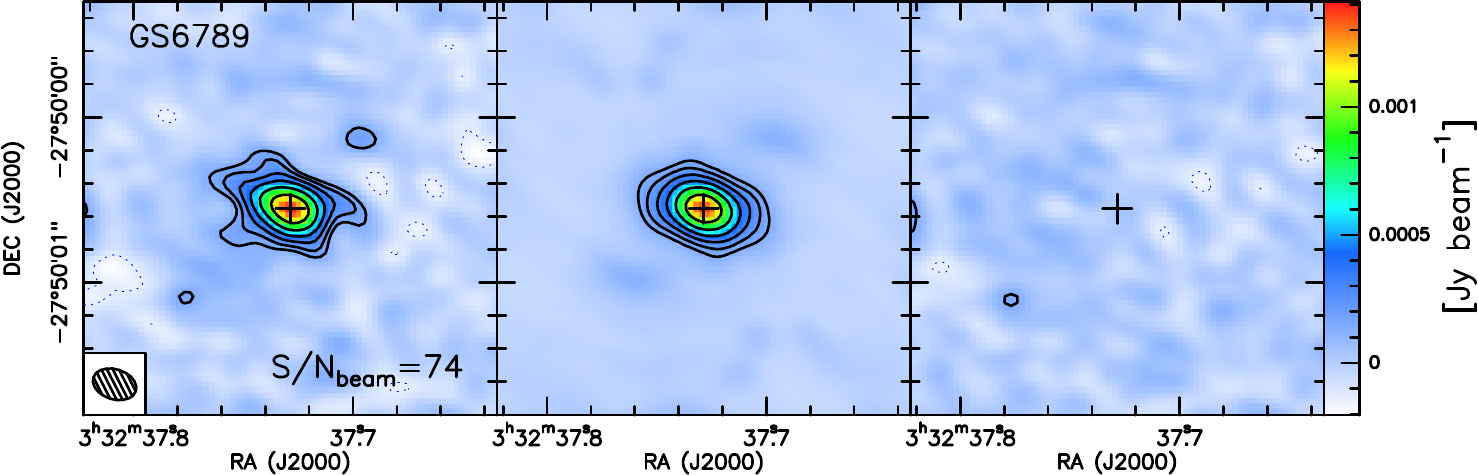}
\includegraphics[width=0.28\linewidth]{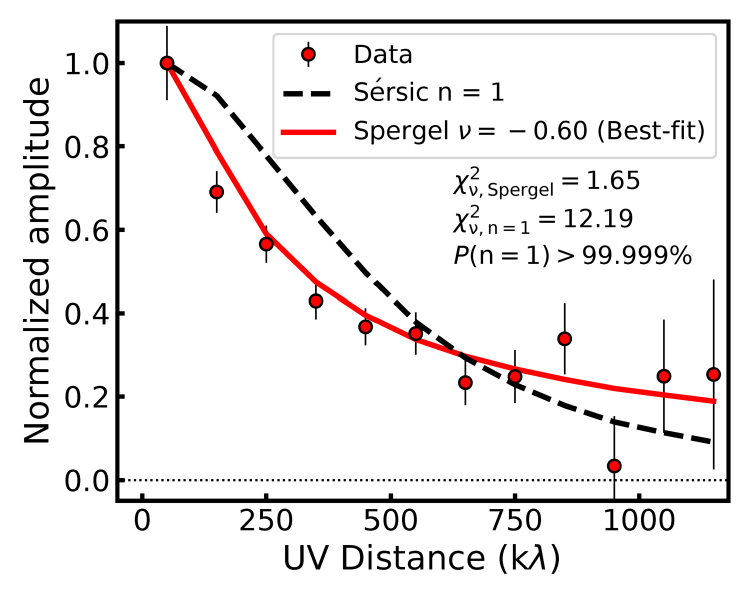}
\captionsetup{labelfont=bf,name=Extended Data Fig.,labelsep=period}
\caption{\textbf{Examples of the best-fit results of the Spergel profile fitting for our sample of galaxies on the $uv$-plane.} From left to right, we show the dirty image (natural weighting), source model convolved with the dirty beam, residuals after subtracting the source model, and the normalized visibility amplitudes as a function of $uv$-distance. Contours start at $\pm$3$\sigma$ and increase by a factor of 1.5. The black crosses mark the centers of submillimeter emission from sources derived from Spergel model fitting, while the source name and the S/N$_{\rm beam}$ of the data are indicated in left panels. The ALMA beam is shown in the bottom left corner of the left panel. Red solid curves in the right panels represent the best-fit of Spergel modeling to the $uv$-data. For comparison, an exponential (\sersic \ $n=1$) model is overlaid by a black dashed curve. Error bars show the statistical noise on the average amplitude in each bin. The reduced chi-squared values calculated for the best-fit of Spergel modeling ($\chi^2_{\rm \nu, Spergel}$) and $n=1$ profile fit ($\chi^2_{\rm \nu, n=1}$), along with a degree of confidence ($P(n=1)$) by which the $n=1$ solution can be rejected, are reported in each panel.}
\label{fig:ext-images}
\end{figure*}

\begin{figure*}[htbp]%
\centering
\includegraphics[width=0.45\linewidth]{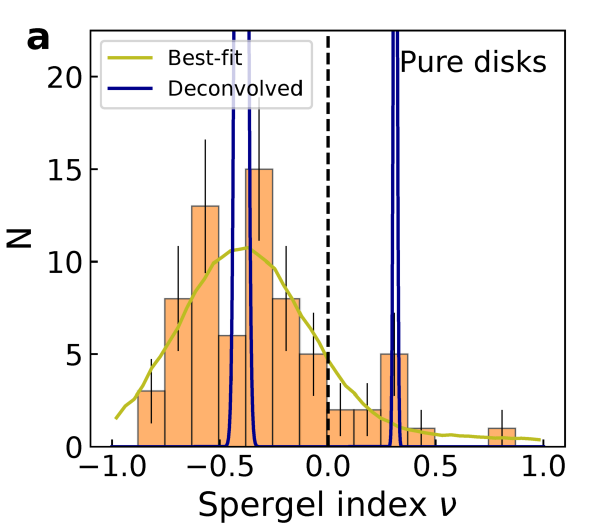}
\includegraphics[width=0.45\linewidth]{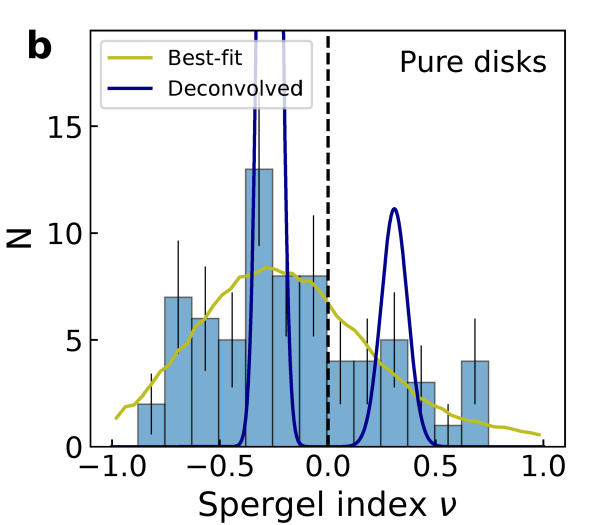}
\captionsetup{labelfont=bf,name=Extended Data Fig.,labelsep=period}
\caption{\textbf{Histogram showing the distribution of Spergel index $\nu$ measured for subsamples of galaxies.} \textbf{a}, $\Sigma_{\rm SFR}$-compact galaxies. \textbf{b}, $\Sigma_{\rm SFR}$-extended galaxies. The blue and olive curves represent the intrinsic and best-fit distributions of $\nu$, respectively. The vertical dashed lines indicate the $\nu=0$ threshold above which the galaxies are classified as pure disks. The error bar in each bin corresponds to the 1$\sigma$ Poisson error.}
\label{fig:ext-nu-distribution}
\end{figure*}

\begin{figure*}[htbp]%
\centering
\includegraphics[width=0.45\linewidth]{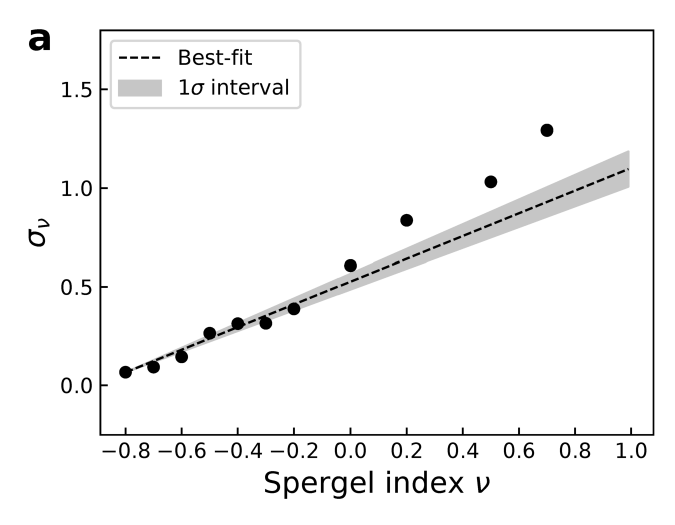}
\includegraphics[width=0.45\linewidth]{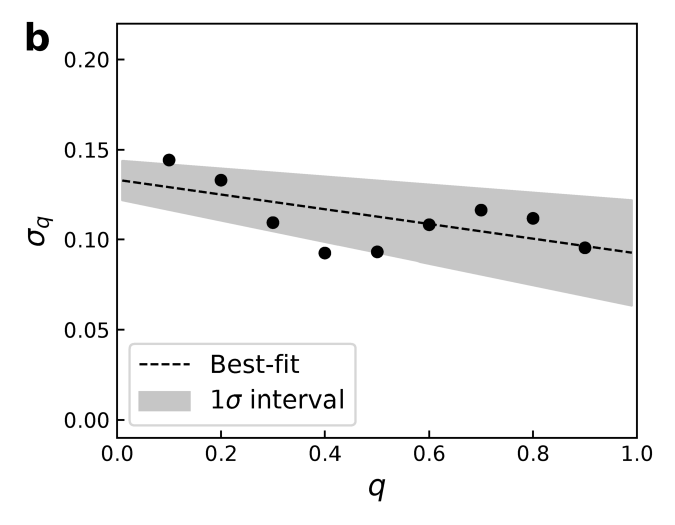}
\captionsetup{labelfont=bf,name=Extended Data Fig.,labelsep=period}
\caption{\textbf{Relative precision in the measurement of Spergel index $\nu$ and apparent axis ratio \textit{q}}. \textbf{a}, the uncertainty $\sigma_\nu$ is a measure of the accuracy of the recovered $\nu$ in MC simulation, evaluated as the median absolute deviation of the data around the true value ($\sigma = 1.48\times$MAD). The dashed line represents the best-fit linear relationship between $\sigma_\nu$ and $\nu$ and the shaded region indicates the 1$\sigma$ confidence interval. \textbf{b}, similar to panel \textbf{a} but showing the uncertainty in $q$ measurement. }
\label{fig:ext-sigma-function}
\end{figure*}

\begin{figure*}[htbp]%
\centering
\includegraphics[width=0.4\linewidth]{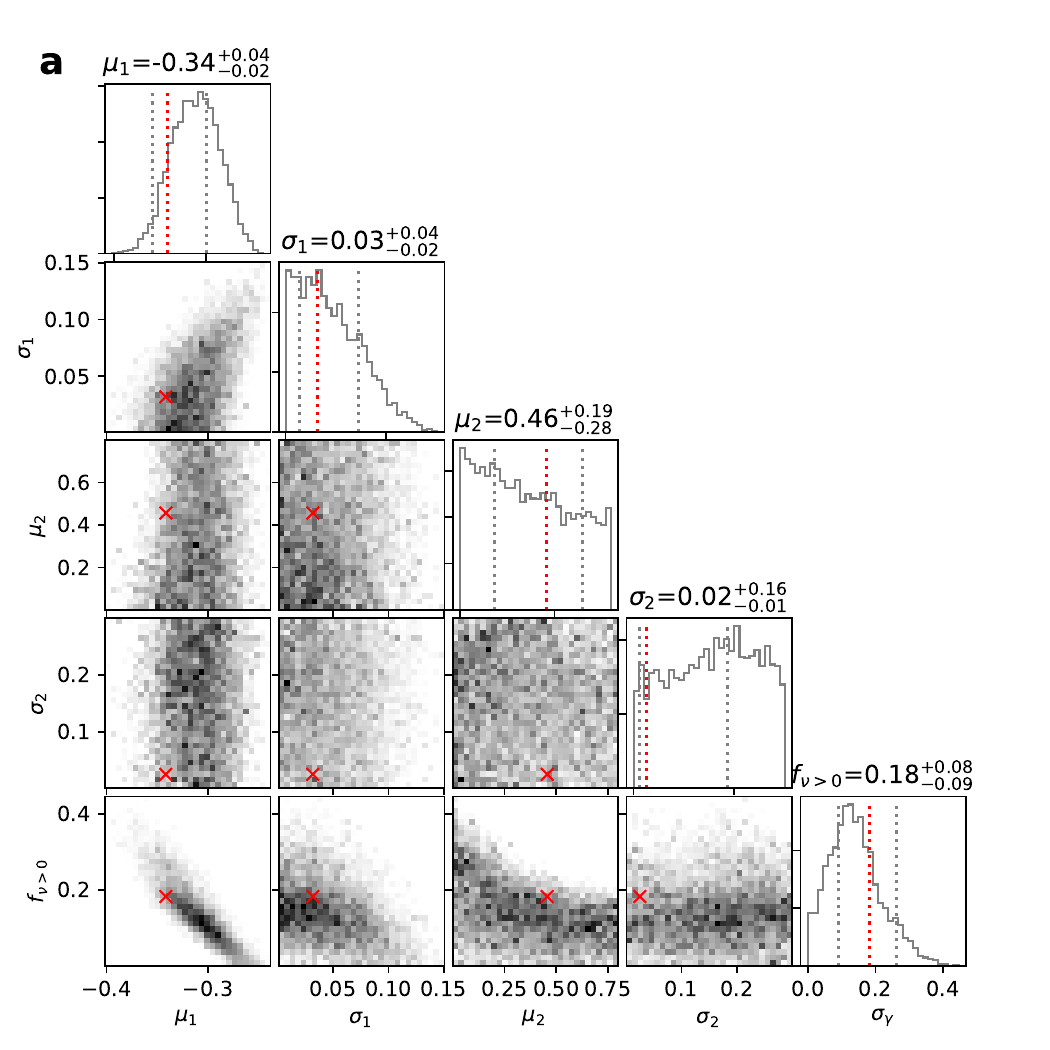}
\includegraphics[width=0.35\linewidth]{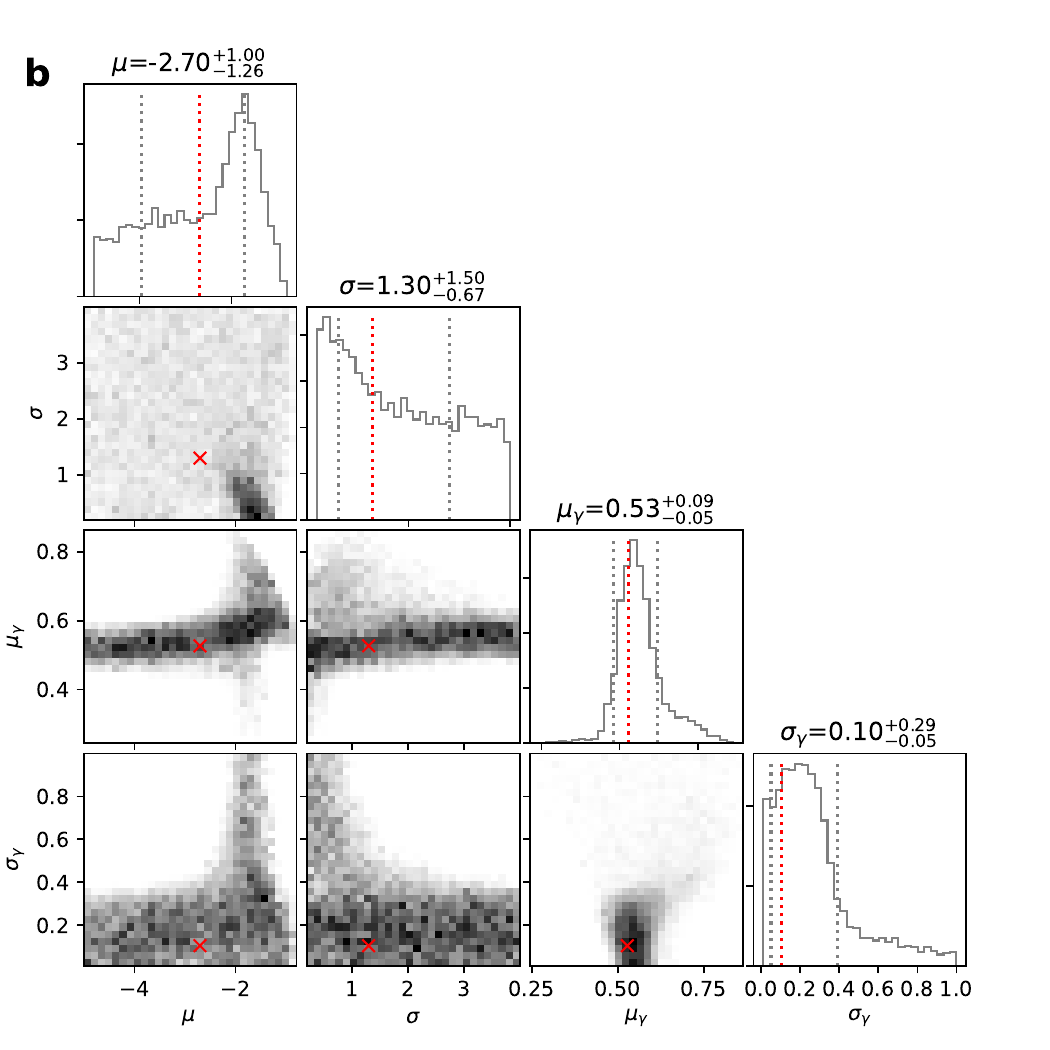}\\
\includegraphics[width=0.4\linewidth]{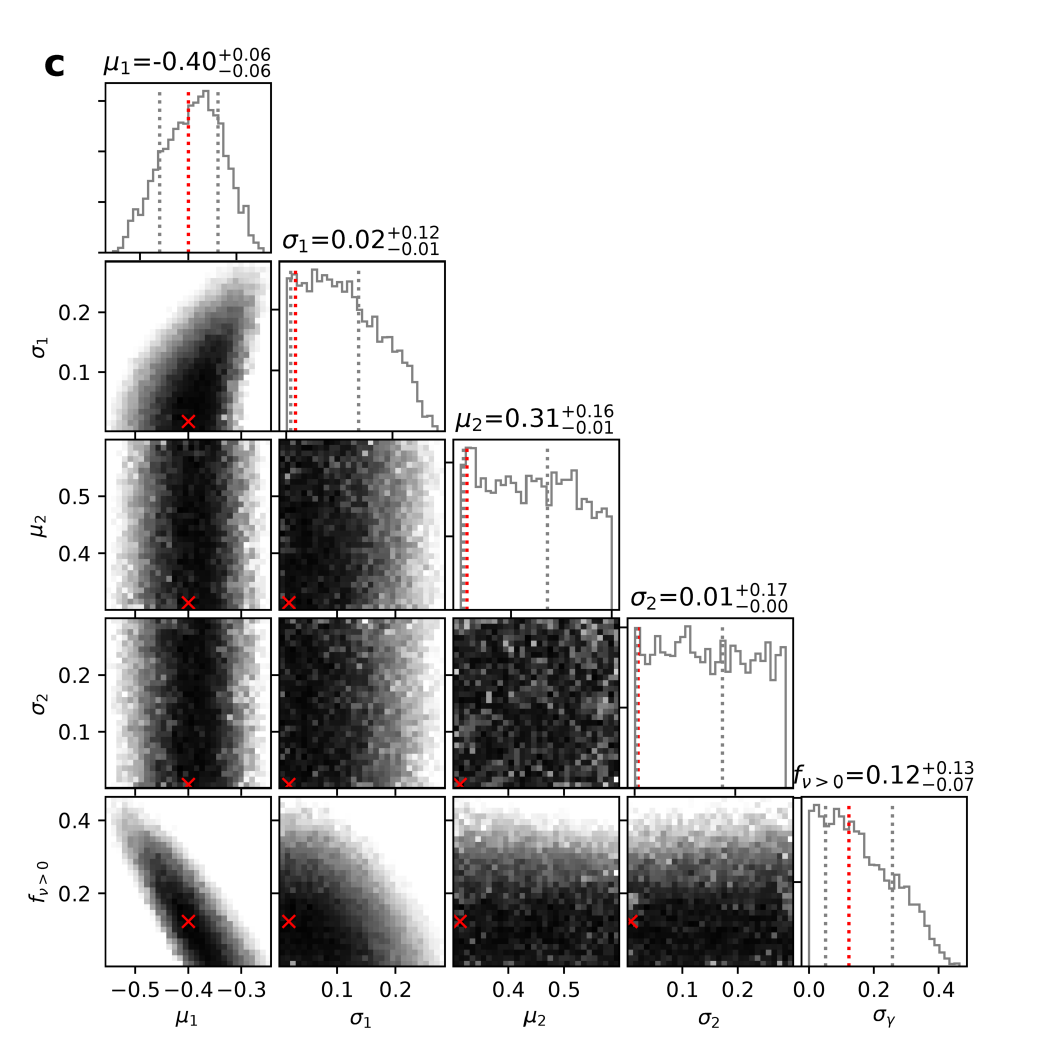}
\includegraphics[width=0.35\linewidth]{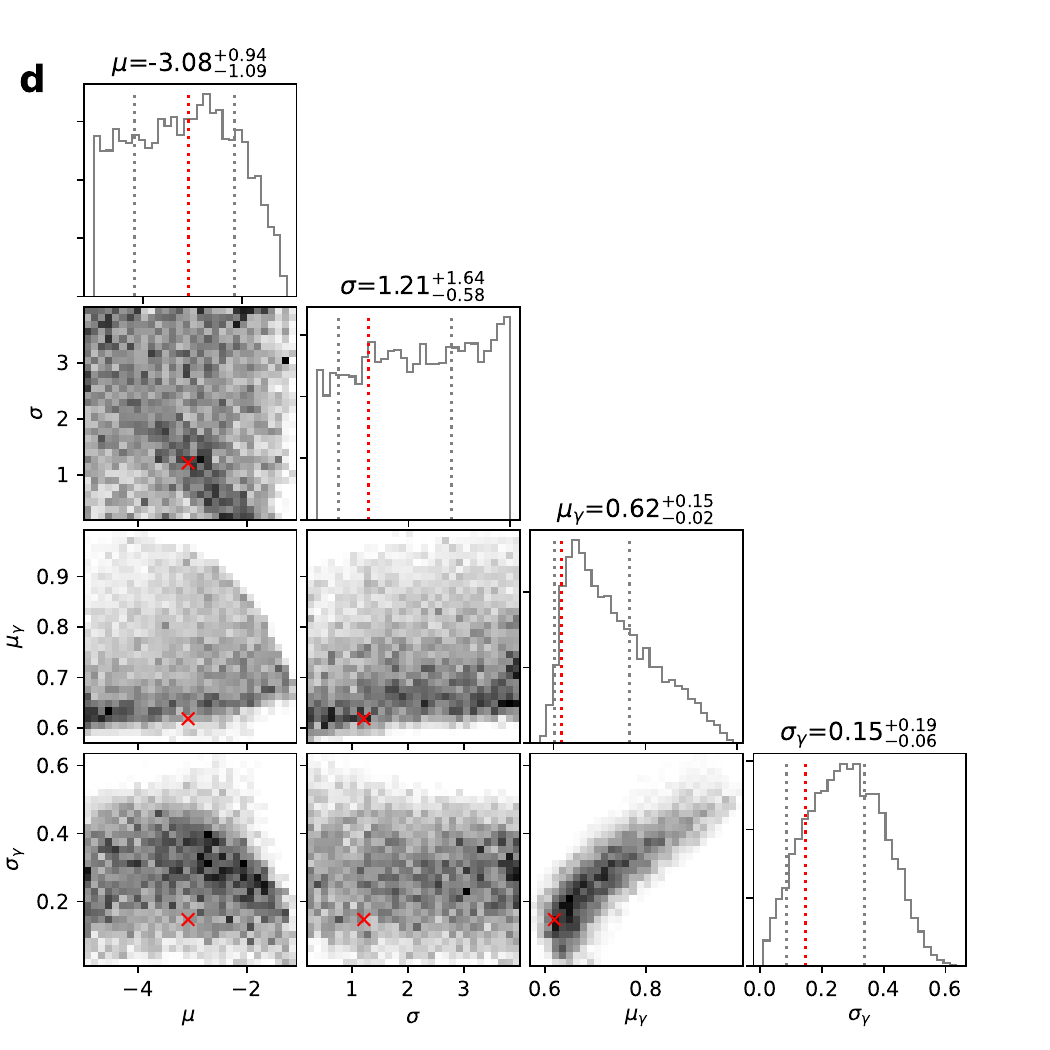}\\
\includegraphics[width=0.4\linewidth]{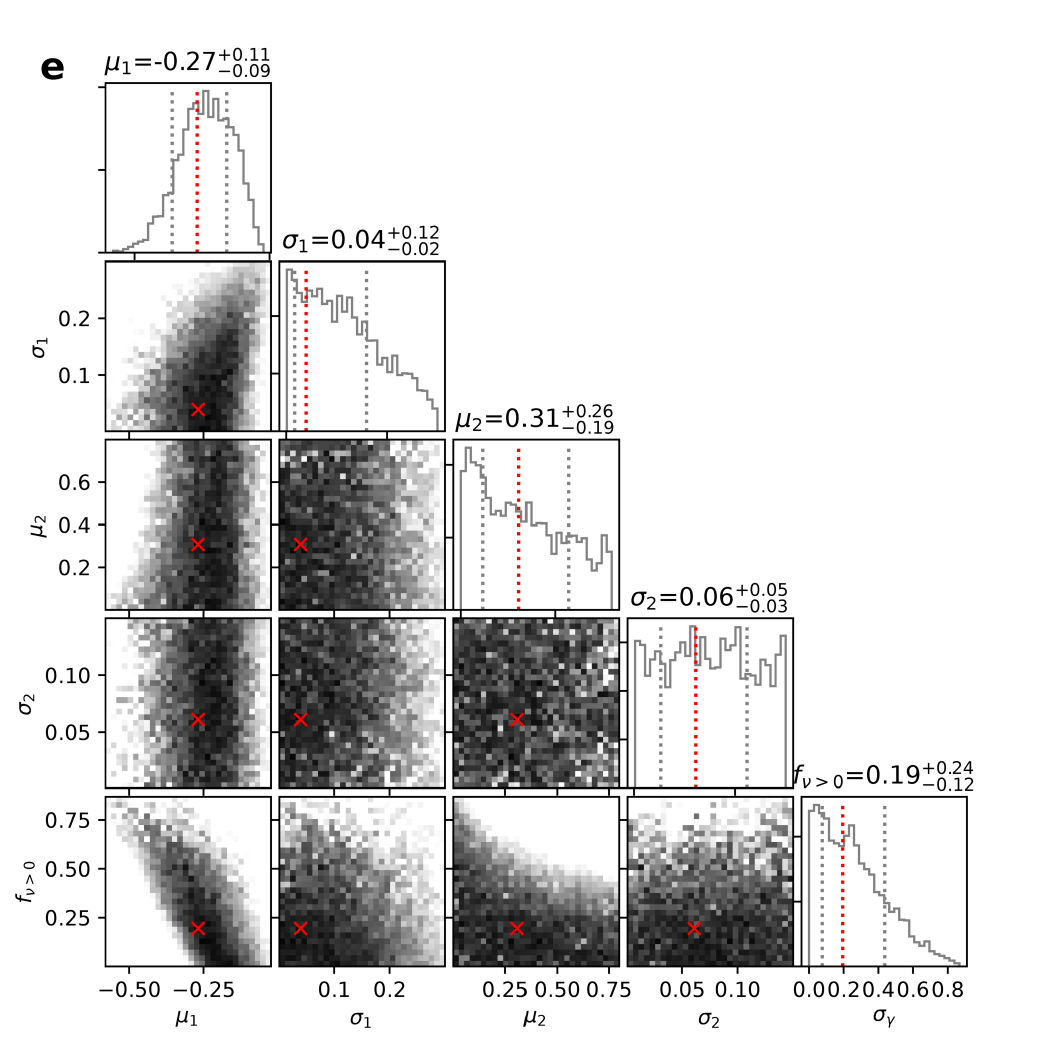}
\includegraphics[width=0.35\linewidth]{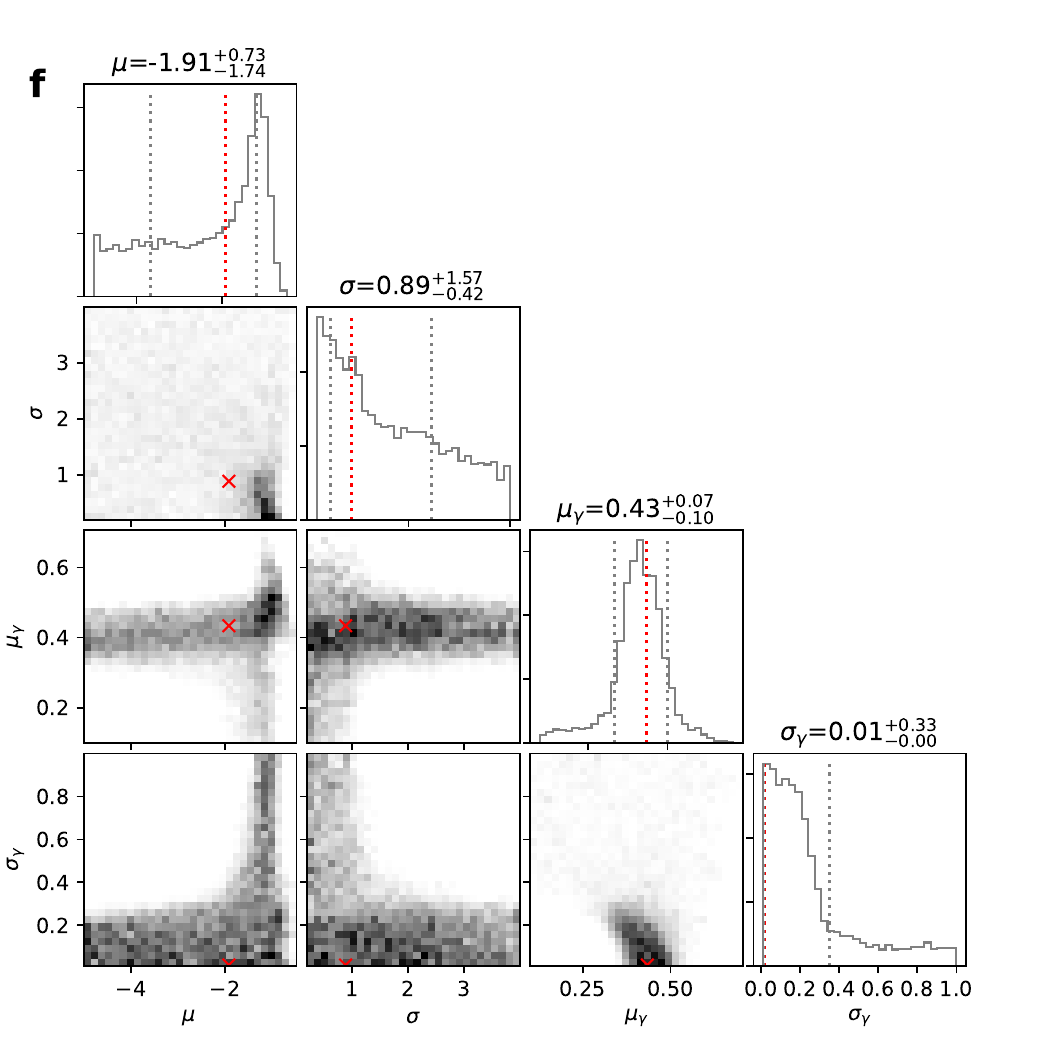}
\captionsetup{labelfont=bf,name=Extended Data Fig.,labelsep=period}
\caption{\textbf{Corner plots showing the projections of the posterior probability distributions of the fitted parameters using an MCMC analysis.} \textbf{Left}: probability distribution of the parameters in a two-Gaussian model of the Spergel index $\nu$ distribution. \textbf{Right}: geometric parameter estimation for the triaxial model. \textbf{a,b}, the full sample.  \textbf{c,d}, a subsample of $\Sigma_{\rm SFR}$-compact galaxies. \textbf{e,f}, a subsample of $\Sigma_{\rm SFR}$-extended galaxies. The best-fit values are marked as red crosses and red dotted lines, and listed on top of the histograms, with uncertainties computed as 1$\sigma$ standard deviation (dotted grey lines), using the posterior probability as weights. }\label{fig:ext-mcmc-corner}
\end{figure*}

\begin{figure*}[htbp]%
\centering
\includegraphics[width=0.32\linewidth]{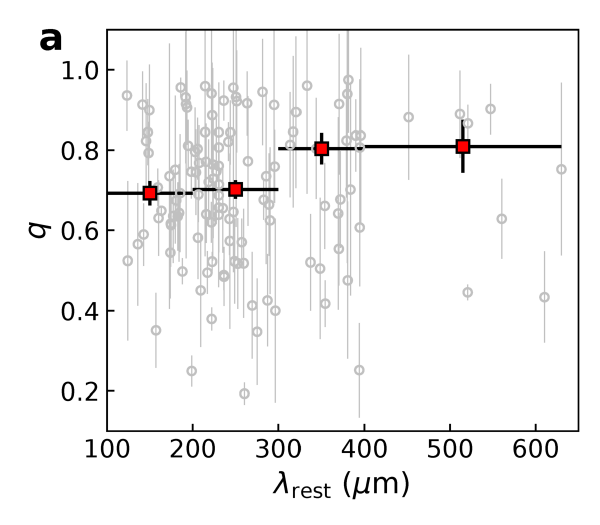}
\includegraphics[width=0.32\linewidth]{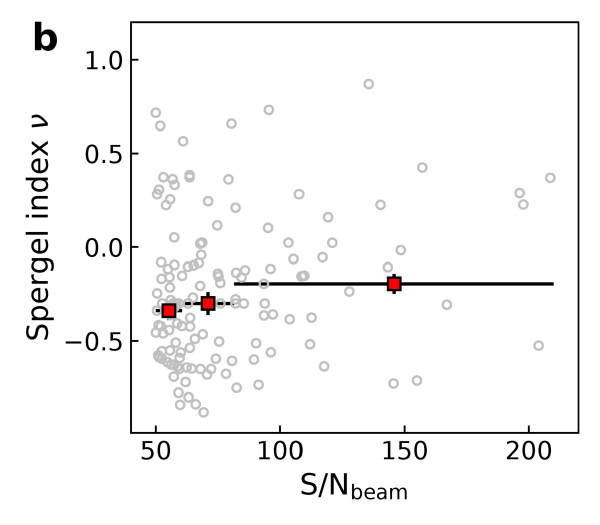}
\includegraphics[width=0.32\linewidth]{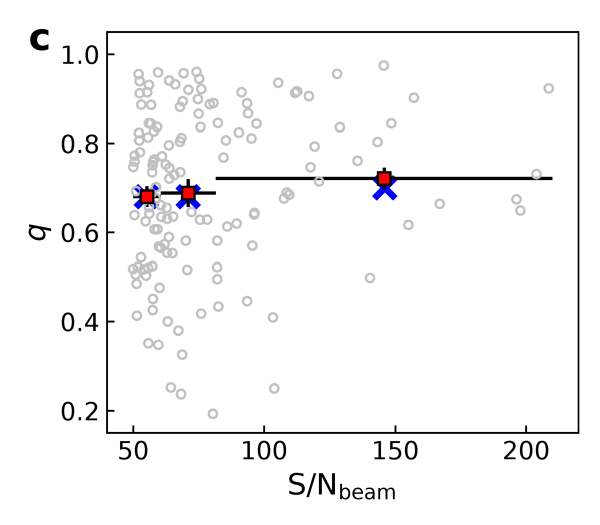}
\captionsetup{labelfont=bf,name=Extended Data Fig.,labelsep=period}
\caption{\textbf{Distribution of measured parameters for the full ALMA sample galaxies.} \textbf{a}, axis ratio $q$ versus rest-frame wavelength $\lambda_{\rm rest}$. \textbf{b}, Spergel \ index $\nu$ versus S/N$_{\rm beam}$. \textbf{c}, $q$ versus S/N$_{\rm beam}$. The red-filled squares in panel \textbf{a} indicate the median values of $q$ in different wavelength bins, while in panel \textbf{b} and \textbf{c} represent the median and mean values of $\nu$ and $q$ in different S/N$_{\rm beam}$ bins, respectively. The vertical and horizontal bars indicate the error on the average and bin width, respectively. The blue crosses in panel \textbf{c} represent the intrinsic distribution of $q$, derived from the best-fit for the whole sample, perturbed by noise. The noise is assumed to be Gaussian with the standard deviation estimated as the error on $q$ measured at the corresponding S/N$_{\rm beam}$ bins, with average values of 0.19, 0.16, and 0.08, respectively. The good agreement between the model and data suggests that the slight decrease in $q$ at lower S/N levels can be attributed to the higher levels of noise observed.}
\label{fig:ext-lambda-snr}
\end{figure*}

\begin{figure*}[htbp]%
\centering
\includegraphics[width=0.24\linewidth]{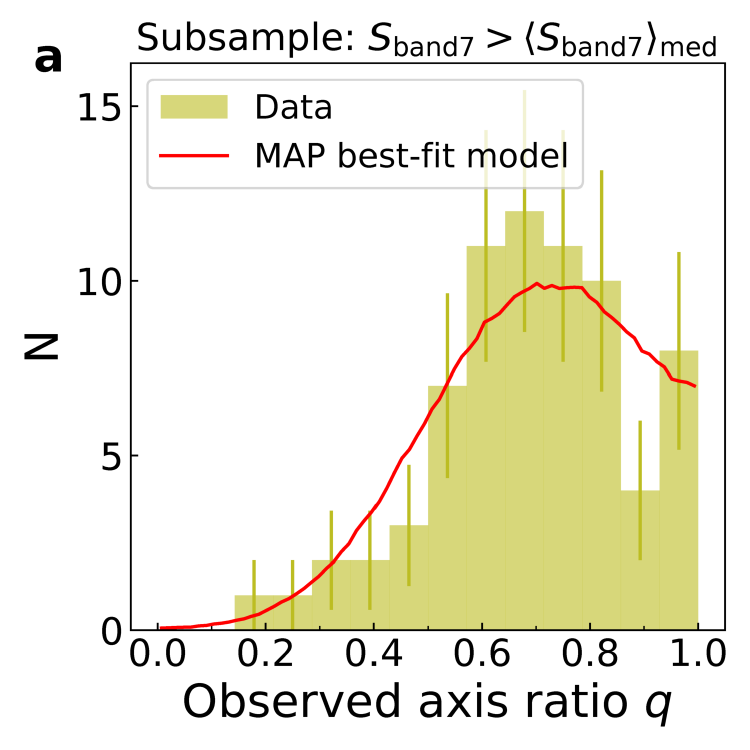}
\includegraphics[width=0.24\linewidth]{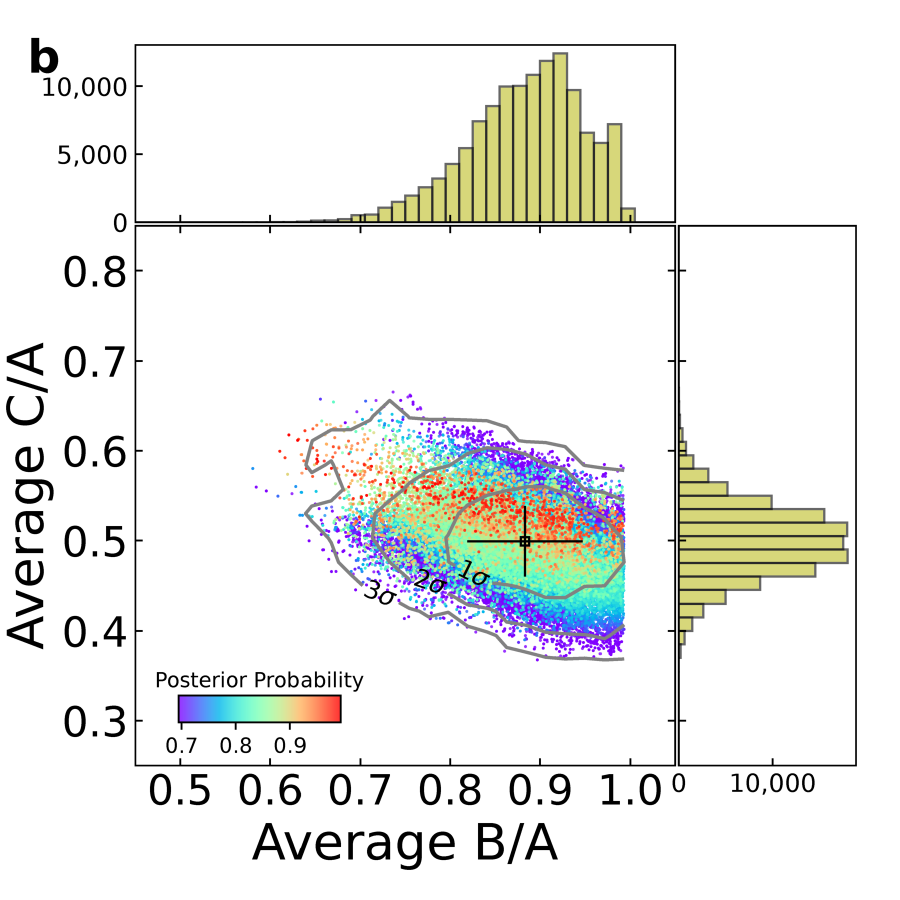}
\includegraphics[width=0.24\linewidth]{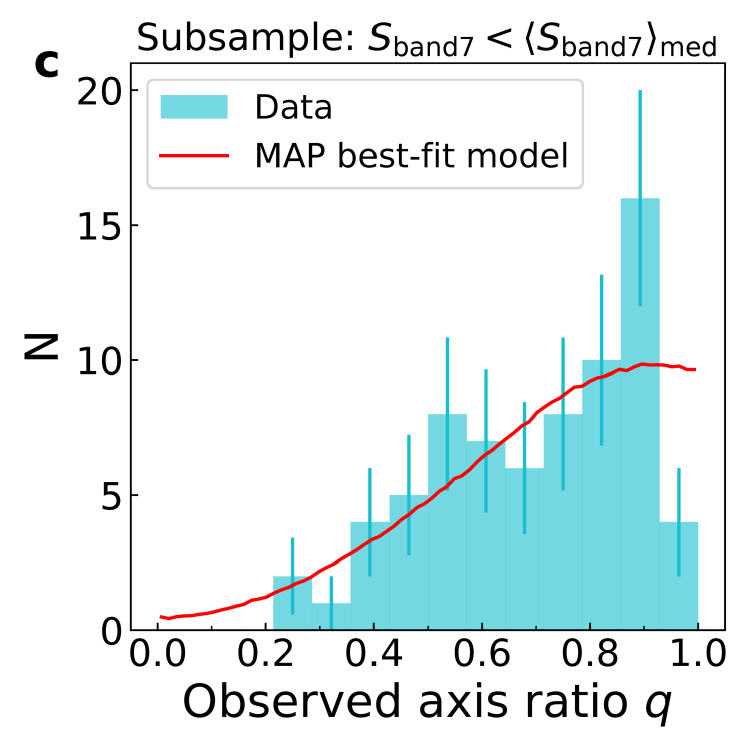}
\includegraphics[width=0.24\linewidth]{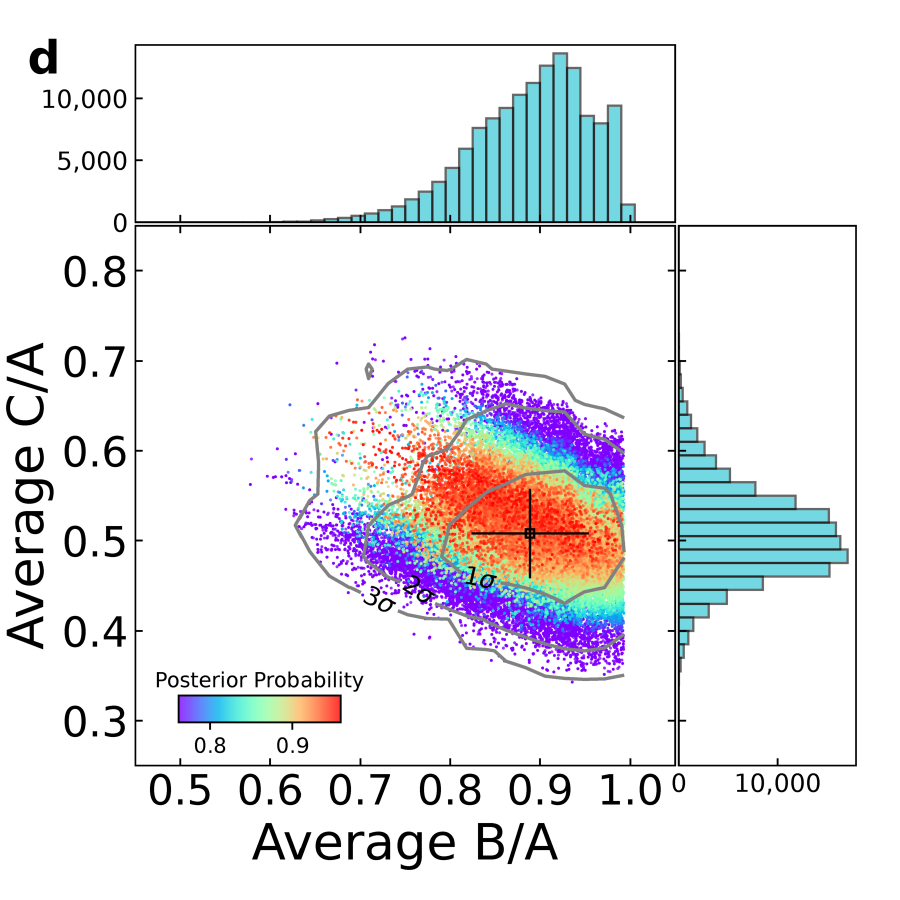}
\includegraphics[width=0.24\linewidth]{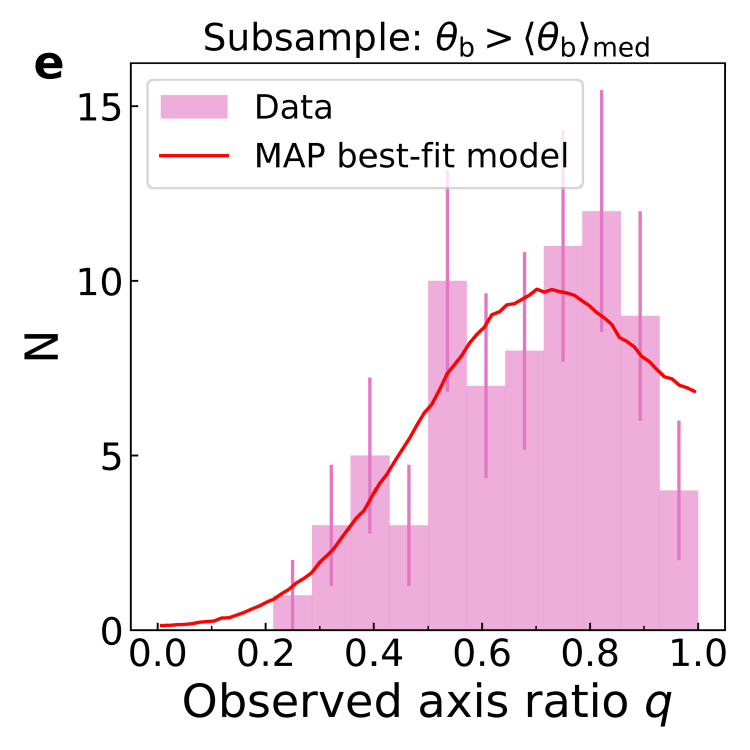}
\includegraphics[width=0.24\linewidth]{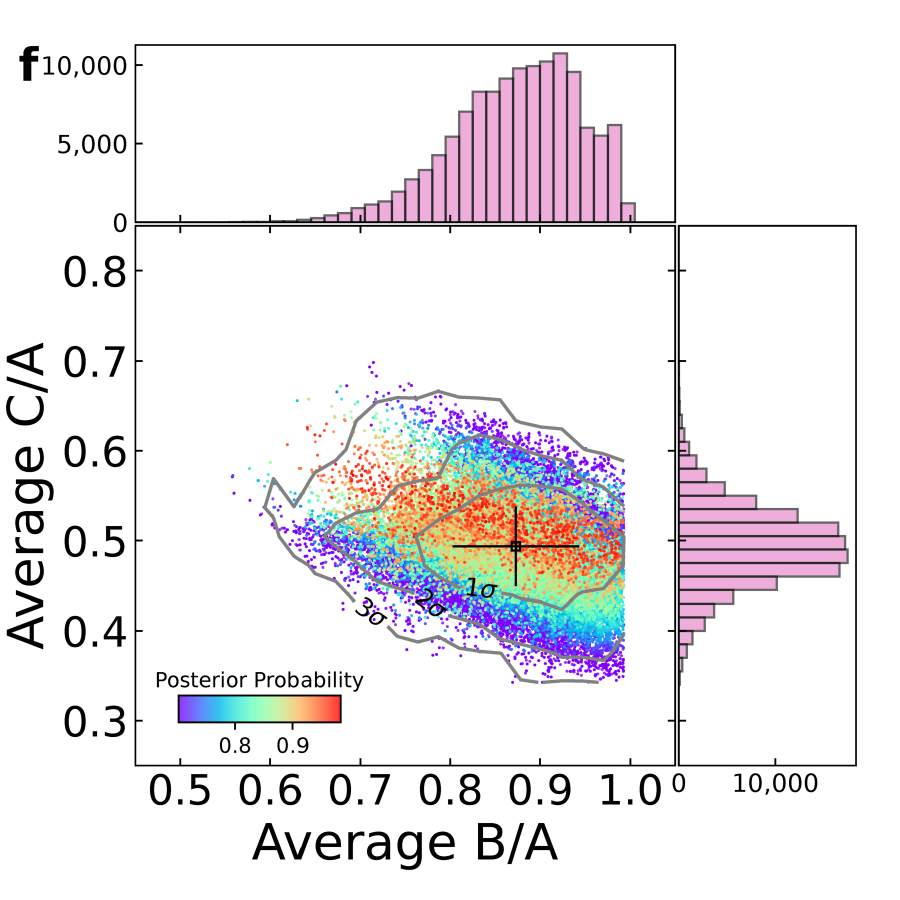}
\includegraphics[width=0.24\linewidth]{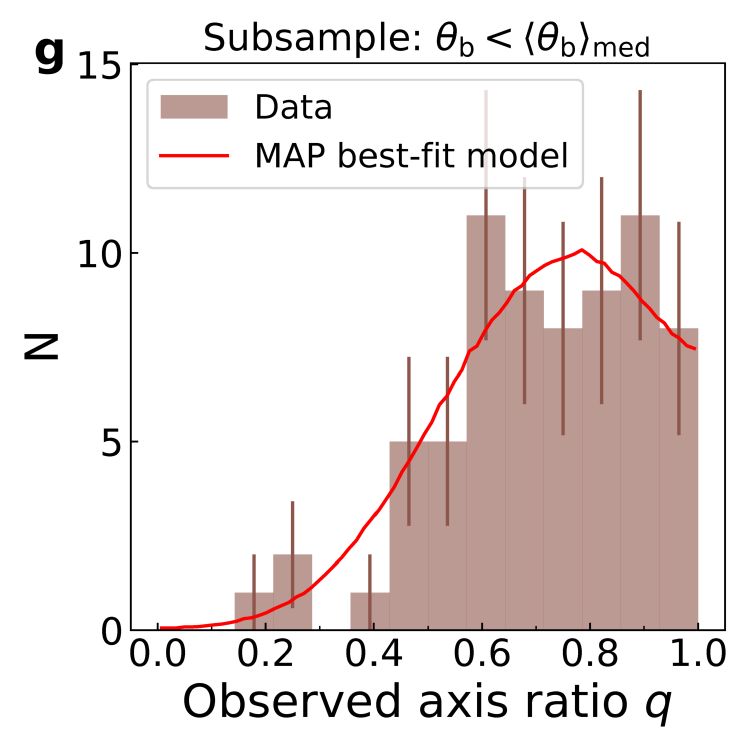}
\includegraphics[width=0.24\linewidth]{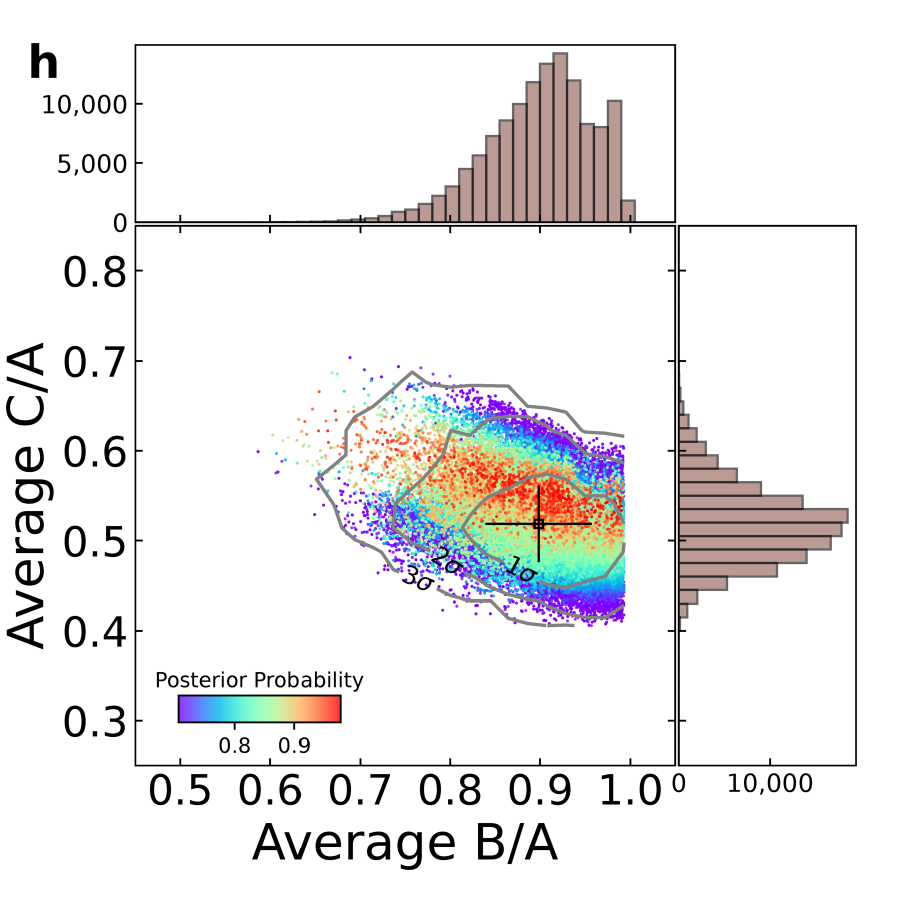}
\captionsetup{labelfont=bf,name=Extended Data Fig.,labelsep=period}
\caption{\textbf{Results from triaxial modeling for the subsamples of galaxies split by flux density and beam size.} The median flux density at band~7 and the median beam size for the full sample are $7.8_{-4.7}^{+3.6}$~mJy and $0.51_{-0.31}^{+0.39}$~arcsec (uncertainties are interquartile range), respectively. \textbf{a,b}, a subsample of galaxies with flux brighter than the median value. The $\langle B/A \rangle$ and $\langle C/A \rangle$ are 0.88$\pm$0.06 and 0.50$\pm$0.04. \textbf{c,d}, a subsample of galaxies with flux fainter than the median value. The $\langle B/A \rangle$ and $\langle C/A \rangle$ are 0.89$\pm$0.07 and 0.51$\pm$0.05. \textbf{e,f}, a subsample of galaxies with beam size larger than the median value. The $\langle B/A \rangle$ and $\langle C/A \rangle$ are 0.87$\pm$0.07 and 0.49$\pm$0.04. \textbf{g,h}, a subsample of galaxies with beam size smaller than the median value. The $\langle B/A \rangle$ and $\langle C/A \rangle$ are 0.90$\pm$0.06 and 0.52$\pm$0.04, respectively.}
\label{fig:ext-subsamples}
\end{figure*}

\begin{figure*}[htbp]%
\centering
\includegraphics[width=0.4\linewidth]{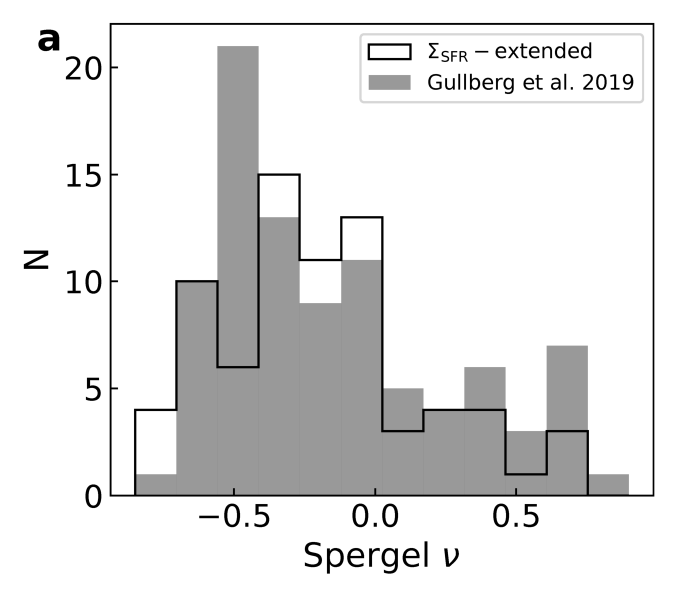}
\includegraphics[width=0.4\linewidth]{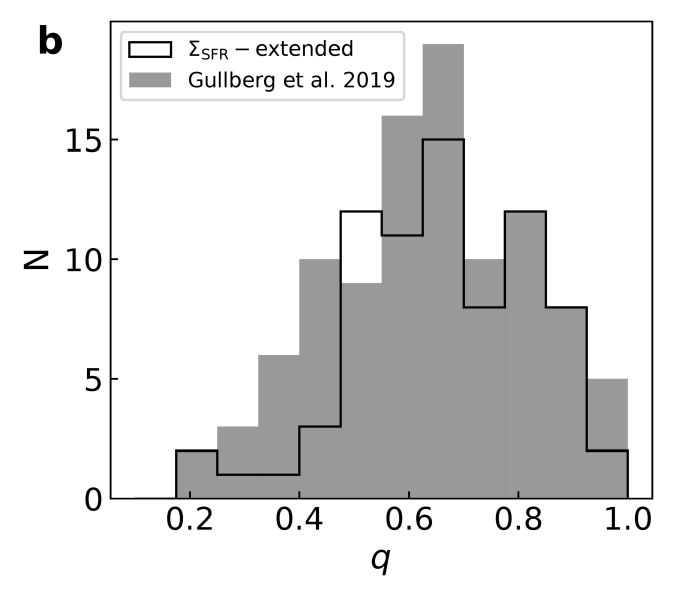}
\includegraphics[width=0.4\linewidth]{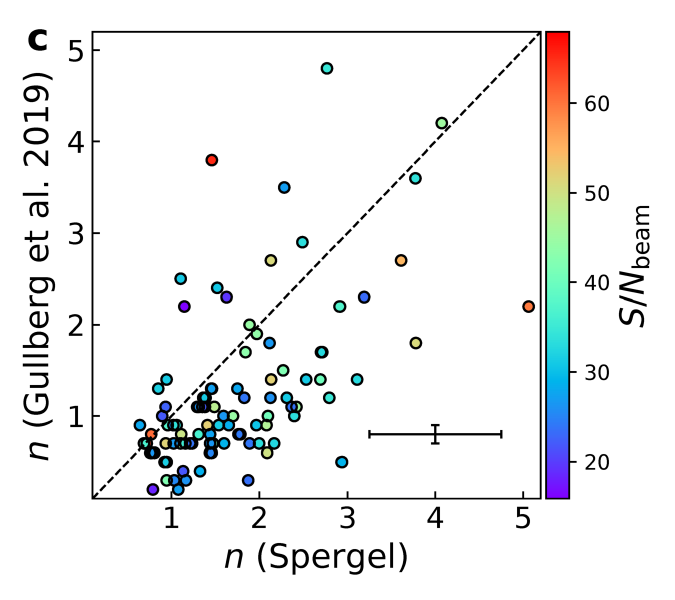}
\includegraphics[width=0.4\linewidth]{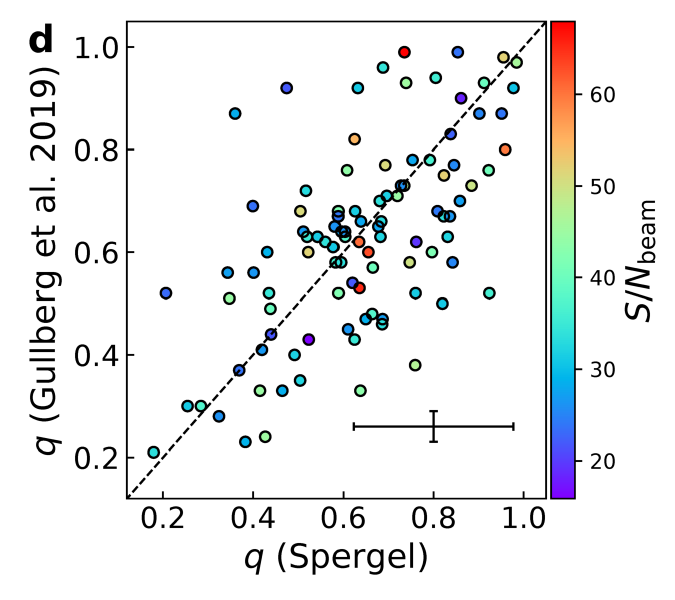}
\includegraphics[width=0.4\linewidth]{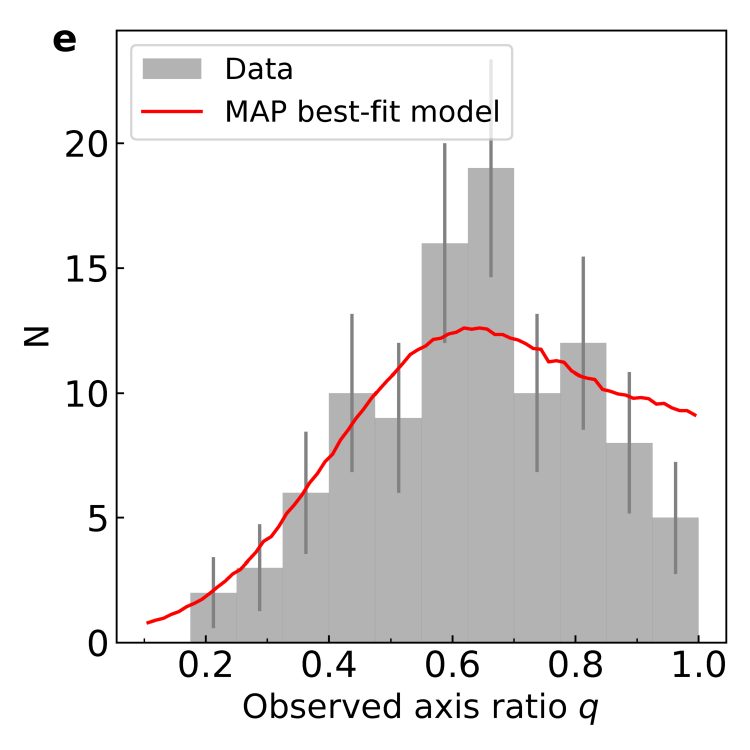}
\includegraphics[width=0.4\linewidth]{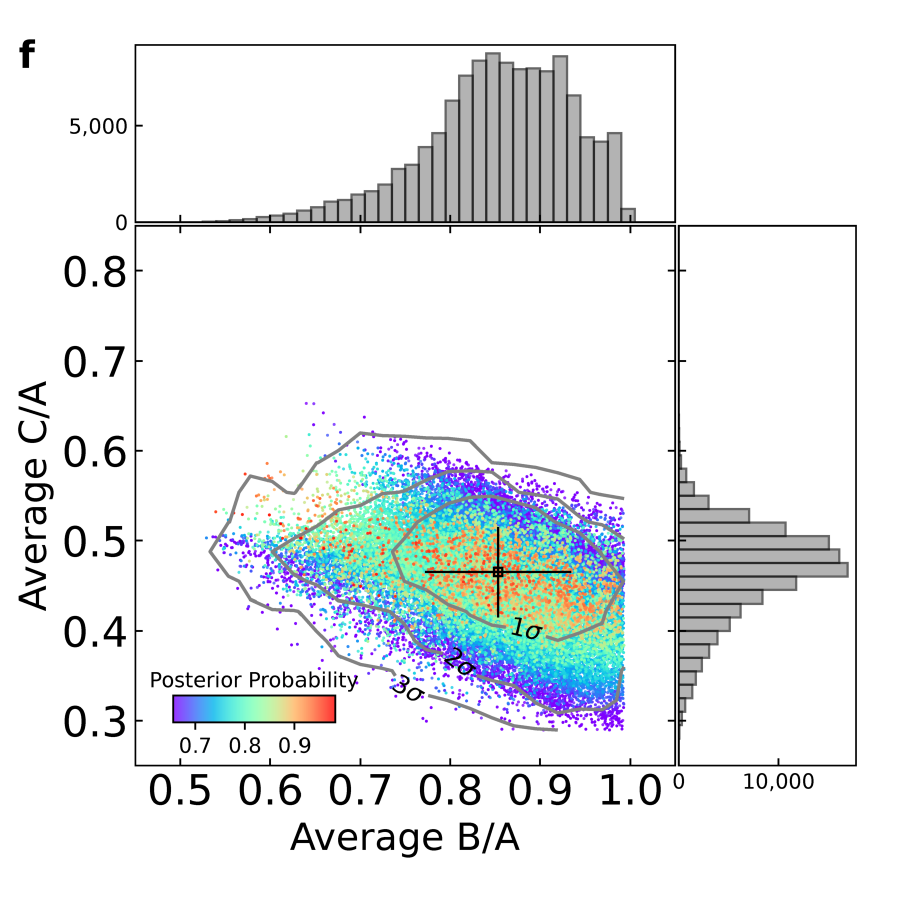}
\captionsetup{labelfont=bf,name=Extended Data Fig.,labelsep=period}
\caption{\textbf{Reanalysis of sample galaxies of SMGs measured with $R_{\rm e}/\Delta R_{\rm e}>3$ in ref.~\cite{Gullberg2019}.} \textbf{a,b}, distribution of Spergel $\nu$ and axis ratio $q$ measured using Spergel modeling for the galaxies in ref.~\cite{Gullberg2019}, compared to the subsample of $\Sigma_{\rm SFR}-$extended galaxies in our sample. \textbf{c,d}, comparison of \sersic \ index $n$ and $q$ measurements between Spergel fitting and those derived from ref.~\cite{Gullberg2019}. Data points are colour-coded by S/N$_{\rm beam}$. The vertical and horizontal bars indicate the median uncertainties derived from Spergel fitting and those reported in ref.~\cite{Gullberg2019}, respectively. The dashed lines indicate the 1:1 line. \textbf{e,f}, similar to Fig.~\ref{fig:hist-axis}, but showing the triaxial modeling results for the sample galaxies in ref.~\cite{Gullberg2019} with parameters fitted using Spergel profile. The measured $\langle B/A \rangle$ and $\langle C/A \rangle$ are 0.85$\pm$0.08 and 0.47$\pm$0.05, respectively.}
\label{fig:ext-gullberg}
\end{figure*}

\begin{figure*}[htbp]%
\centering
\includegraphics[width=0.6\linewidth]{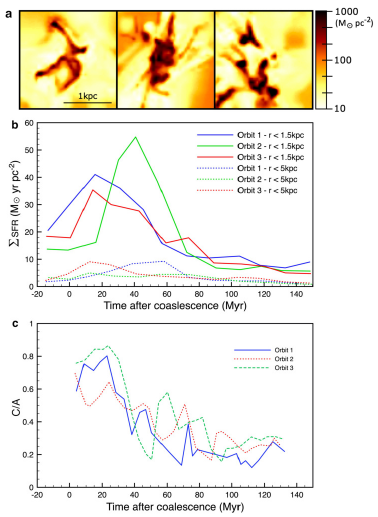}
\captionsetup{labelfont=bf,name=Extended Data Fig.,labelsep=period}
\caption{\textbf{Compact spheroid-like gas distribution from hydro-simulations of major mergers.} Reanalysis of recent very high-resolution simulations of mergers of turbulent clumpy disks \cite{Bournaud2011}. \textbf{a}, maps (2$\times$2 kpc) of the central gas in three different mergers, showing the flattest projection for these systems observed at 12~Myr from coalescence, that is these systems are 3D spheroidal structures, not face-on disks. \textbf{b,c}, evolution of SFR surface density and disk thickness C/A over time after merger coalescence for three merger orbits, respectively. Panel \textbf{b} distinguishes the nuclear regions of the mergers where submm-emission is bright, from the wider outskirts. Time analysis shows that the spheroidal shape of the gas can be maintained over $\sim30$—50~Myr. This is compatible with the inferred timescales for the submm-bright bursts based on observations. 
After the intensely star-forming spheroid-like star formation, lacking further turbulent energy injection back into the system, the residual gas flattens into a disk (panel \textbf{c}). However, the earlier phase led to the formation of a stellar spheroid.
 }
 \label{fig:hydro-simulations}
\end{figure*}

\begin{figure*}[htbp]
\centering
    \includegraphics[angle=0,width=\linewidth]{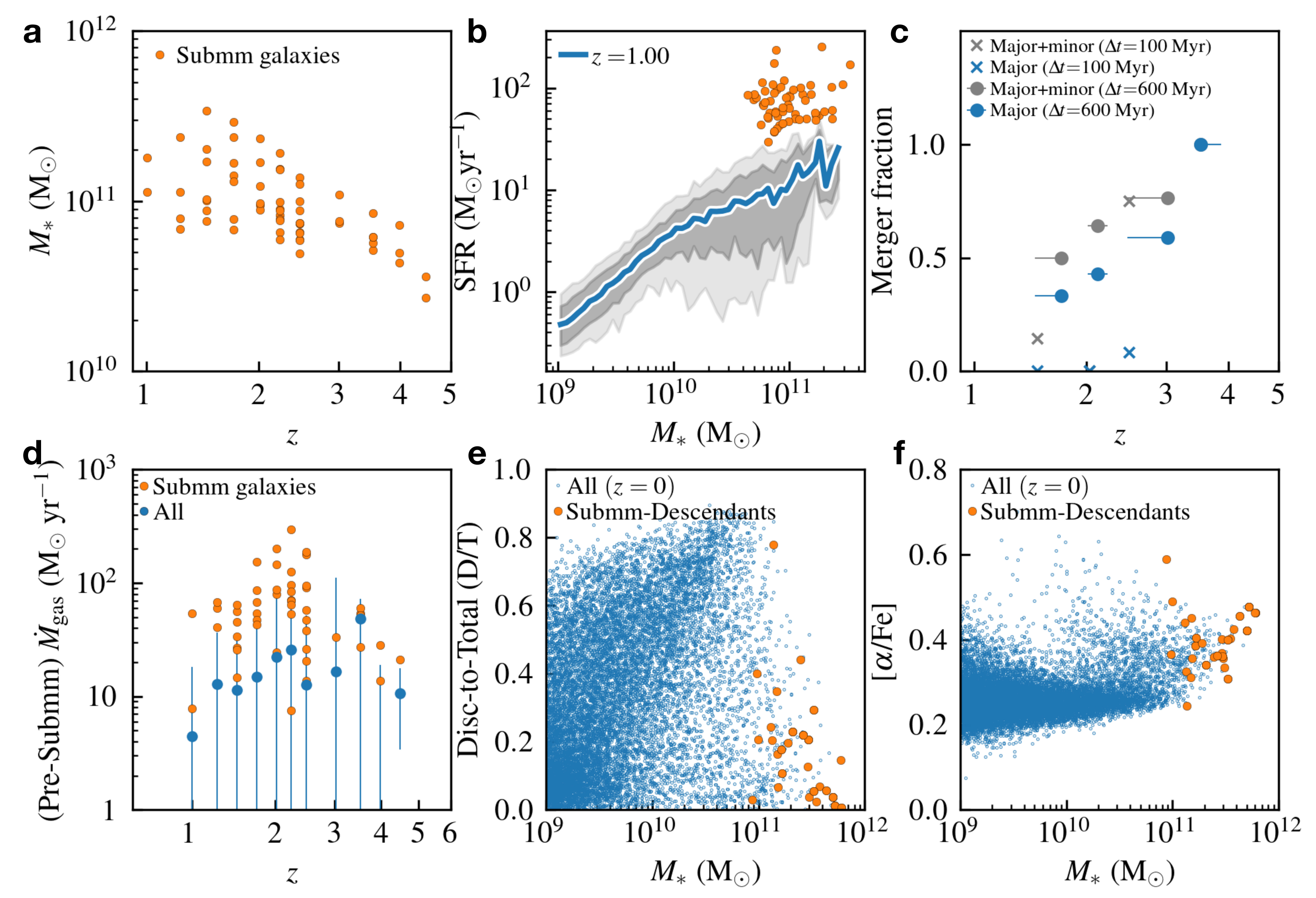}
    \captionsetup{labelfont=bf,name=Extended Data Fig.,labelsep=period}
	\caption{\textbf{Merger history of submm galaxies using EAGLE cosmological simulation.} \textbf{a}, Stellar mass and redshift distribution of the submm galaxy sample (orange symbols). \textbf{b}, SFR of the submm sample (shown in orange symbols) relative to the star-forming Main Sequence of the simulations  (blue solid line; all trends rescaled in star-formation to $z=1$, for clarity). Dark grey shaded regions mark the 16-84th percentiles of the relation, whereas light grey shaded regions indicate the 5-95th percentiles. Individual SFRs of the submm sample have been renormalized (see text). \textbf{c}, fraction of submm galaxies that underwent a major (blue symbols) or minor (grey symbols) mergers. The panel shows that major mergers do not seem to dominate the assembly of submm-bright galaxies. \textbf{d}, gas accretion rates of galaxies before they became submm bright. The accretion includes diffuse gas and minor and (rare) major mergers. Blue symbols indicate the median rates of a control sample, consisting of galaxies with stellar masses within $\pm 0.2$ dex of the median stellar mass of the submm galaxies. This panel indicates that the submm event was likely triggered by a high rate of gas inflow. \textbf{e}, Disc-to-Total (D/T) mass ratio of all galaxies at $z=0$. Orange symbols highlight the D/T ratio of the $z=0$ descendants of the submm galaxies. \textbf{f}, Similar to panel \textbf{e}, stellar [$\alpha$/Fe] (represented by [O/Fe]) of all galaxies at $z=0$, as well as of the submm descendants. Panels \textbf{e} and \textbf{f} indicate that submm galaxies evolve into elliptical galaxies with typical [$\alpha$/Fe] element ratios.}
	\label{Eagle_data}
\end{figure*}

\setcounter{figure}{0}

\begin{figure*}[h]%
\centering
\includegraphics[width=0.8\textwidth]{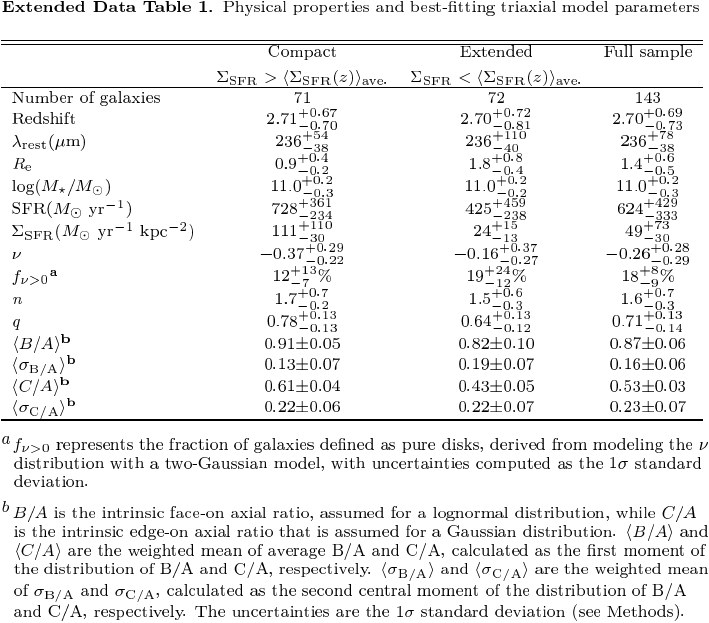}
\captionsetup{labelformat=empty}
\caption{}
\label{tab:properties}
\end{figure*}

\begin{figure*}[h]%
\centering
\includegraphics[width=\textwidth]{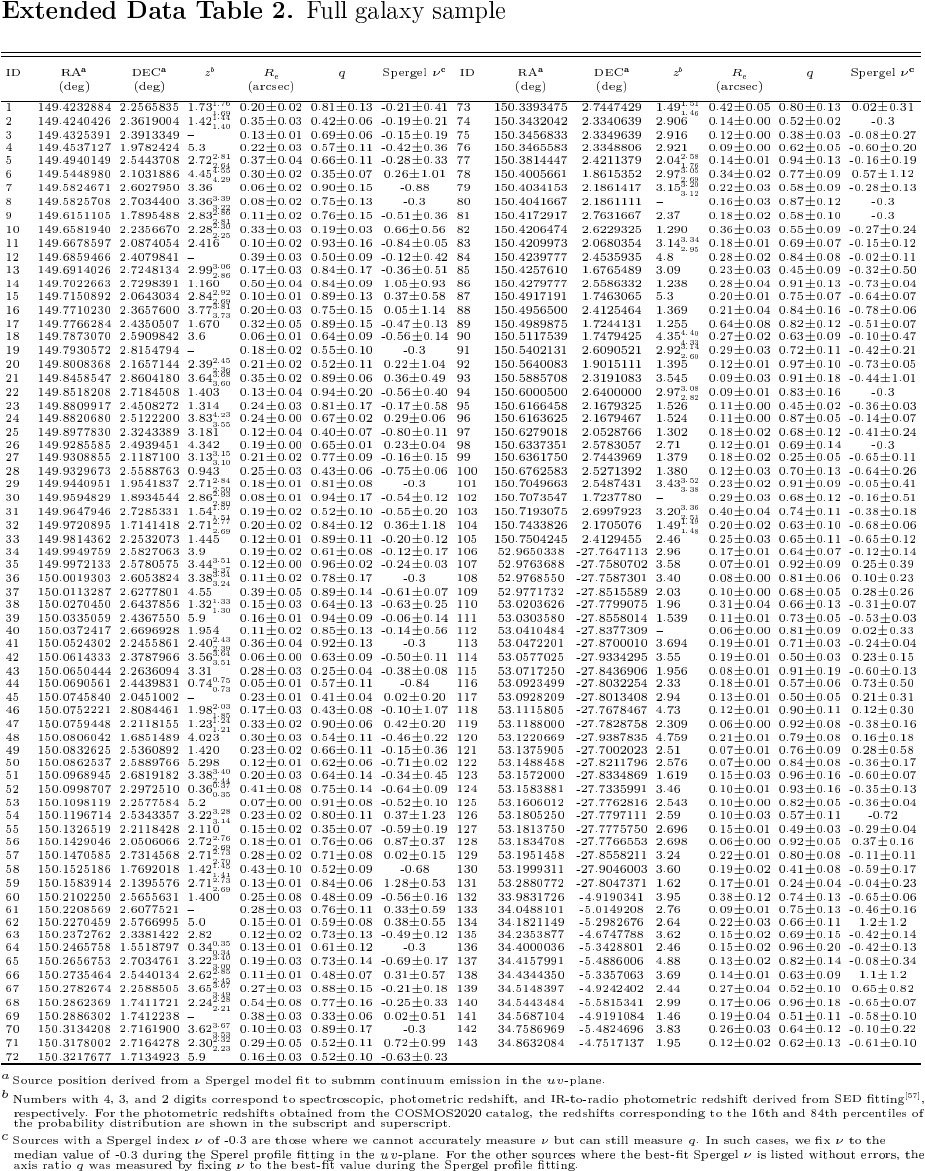}
\captionsetup{labelformat=empty}
\caption{}
\label{tab:measurements}
\end{figure*}

\end{extdata}

\vspace{3pt}
\noindent\rule{\linewidth}{0.4pt}
\vspace{3pt}

\begin{addendum}

\item[Data availability] 

The A$^3$COSMOS submm imaging data are publicly available at \url{https://sites.google.com/view/a3cosmos}. The COSMOS2020 catalogs are available from the Institut d'Astrophysique de Paris (\url{https://cosmos2020.calet.org}), and the COSMOS super-deblended photometry catalog is published at \url{https://cdsarc.cds.unistra.fr/viz-bin/cat/J/ApJ/864/56}.

\item[Code availability] 

The ALMA submm data were reduced using CASA (\url{https://casa.nrao.edu}) and the MAPPING procedure of GILDAS (\url{https://iram.fr/IRAMFR/GILDAS/}). 

\item[Acknowledgements]

QT acknowledges the support by China Scholarship Council (CSC) and the NSFC grant No. 12033004. SA gratefully acknowledges the Collaborative Research Center 1601 (SFB 1601 
sub-project C2) funded by the Deutsche Forschungsgemeinschaft (DFG, German 
Research Foundation) – 500700252. DL acknowledges the support from the Strategic Priority Research Program of the Chinese Academy of Sciences, grant No. XDB0800401.

\item[Authors' contributions]

QT and ED developed the initial idea and led the analysis and the writing of this manuscript. 
BM, SA, and DL led processing of the ALMA data. FB and CAC led the hydrodynamical simulations and the analysis of EAGLE simulations of galaxy mergers, respectively. SZ contributed to the computation of triaxial modeling.
All authors contributed to the development of the analysis and/or 
the interpretation of the results.

\item[Competing interests]

The authors declare no competing interests.

\item[Additional information]

Correspondence and requests for materials should be addressed to Qing-Hua Tan and Emanuele Daddi.

\end{addendum}

\vspace{3pt}
\noindent\rule{\linewidth}{0.4pt}
\vspace{3pt}

\bibliography{sn-bibliography}

\appendix

\end{document}